\documentclass[pdflatex,sn-mathphys]{sn-jnl}

\jyear{2022}%

\theoremstyle{thmstyleone}%
\theoremstyle{thmstyletwo}%

\theoremstyle{thmstylethree}%

\begin{document}

\title[Temperature inhomogeneities cause the abundance discrepancy in H\,II regions]{Temperature inhomogeneities cause the abundance discrepancy in H\,II regions}


\author*[1]{\fnm{J. Eduardo} \sur{M\'endez-Delgado}}\email{jemd@uni-heidelberg.de}


\author[2,3]{\fnm{C\'esar} \sur{Esteban}}

\author[2,3]{\fnm{Jorge} \sur{Garc\'ia-Rojas}}

\author[1]{\fnm{Kathryn} \sur{Kreckel}}

\author[4]{\fnm{Manuel} \sur{Peimbert}}

\affil*[1]{\orgdiv{Astronomisches Rechen-Institut}, \orgname{Zentrum f\"ur Astronomie der Universit\"at Heidelberg}, \orgaddress{\street{M\"onchhofstraße 12-14}, \city{Heidelberg}, \postcode{D-69120}, \state{Baden-W\"urttemberg}, \country{Germany}}}


\affil[2]{\orgdiv{Instituto de Astrof\'isica de Canarias}, \orgname{(IAC)}, \orgaddress{\street{V\'ia L\'actea, 1}, \city{San Crist\'obal de La Laguna}, \postcode{E-38205}, \state{Santa Cruz de Tenerife}, \country{Spain}}}

\affil[3]{\orgdiv{Departamento de Astrof\'isica}, \orgname{Universidad de La Laguna}, \orgaddress{\street{Astrof\'isico Francisco S\'anchez, s/n.}, \city{San Crist\'obal de La Laguna}, \postcode{E-38206}, \state{Santa Cruz de Tenerife}, \country{Spain}}}

\affil[4]{\orgdiv{Instituto de Astronom\'ia}, \orgname{Universidad Nacional Aut\'onoma de M\'exico}, \orgaddress{\street{Apartado Postal 70-264}, \city{Coyoac\'an}, \postcode{04510}, \state{Mexico City}, \country{M\'exico}}}


\abstract{
H\,II regions, ionized nebulae where massive star formation has taken place, exhibit a wealth of emission lines that are the fundamental basis for estimating the chemical composition of the Universe. For more than 80 years, a discrepancy of at least a factor of two between heavy-element abundances derived with collisional excited lines (CELs) and the weaker recombination lines (RLs) has thrown our absolute abundance determinations into doubt. Heavy elements regulate the cooling of the interstellar gas, being essential to the understanding of several phenomena such as nucleosynthesis, star formation and chemical evolution. In this work, we use the best available deep optical spectra of ionized nebulae to analyze the cause of this abundance discrepancy problem. We find for the first time general observational evidence in favor of the temperature inhomogeneities within the gas, quantified by $t^2$. The temperature inhomogeneities inside H\,II regions are affecting only the gas of high ionization degree and producing the abundance discrepancy problem. This work implies that the metallicity determinations based on CELs must be revised, as they can be severely underestimated, especially in the regions of lower metallicity, such as the JWST high-z galaxies. We present methods to estimate these corrections, which will be critical for robust interpretations of the chemical composition of the Universe over cosmic time. 
}

\keywords{H\,II regions, Interstellar abundances, Galaxy abundances, Galaxy chemical evolution}



\maketitle


Since the pioneering work by \cite{Wyse1942}, in essentially all studies where the weak recombination lines (RLs) of heavy-element ions have been detected in ionized nebulae, they imply systematically higher heavy element abundances than those derived from the stronger collisionally excited lines (CELs). Given that the line emission is produced by the same ion through different mechanisms, the abundance discrepancy points to inconsistencies or missing physics in the assumptions underlying the derivation of chemical abundances. Several hypotheses have been proposed over the last 80 years in attempts to resolve this issue. For instance, temperature inhomogeneities \cite{Peimbert1967}, chemical inhomogeneities \cite{Torres-Peimbert1990}, errors in the atomic recombination coefficients \cite{rodriguez2010}, a non-Maxwellian electron velocity distribution \cite{Nicholls2012}, recombination contributions to the auroral CELs \cite{Rubin1986} and some others \cite{Peimbert2017,garciarojas20}. The lack of clear observational confirmation of any of these ideas has prevented a general solution to this problem until now.


In H\,II regions the highly ionized gas is usually concentrated in the volume closest to the massive O or early B type stars, while the less ionized ions are concentrated in outer shells. Therefore, it is common to estimate a representative temperature for each volume of gas using CEL intensity ratios, usually $T_{\rm e}$([O\,III] $\lambda 4363/\lambda 5007$) and $T_{\rm e}$([N\,II] $\lambda 5755/\lambda 6584$) for the volume with high and low degree of ionization, respectively. The reason why the existence of temperature inhomogeneities within a given ionization zone can produce the abundance discrepancy is that the emission of CELs increases exponentially with temperature, whereas the emissivity of RLs depends on this parameter to a much lesser extent (decreasing proportionally to a power of $T_{\rm e}$ close to 1). Therefore, in case of inhomogeneities, $T_{\rm e}$ derived from CEL intensity ratios will be systematically biased towards higher values than the average temperature \cite{Peimbert1967}. Then, by deriving the chemical composition based on the intensity ratio of CELs and H\,I RLs, an exponential systematic error towards lower abundances will be introduced. On the other hand, the intensity ratio of RLs is less sensitive to temperature, making it more suitable for deriving the true chemical composition of the gas. In the paradigm proposed by \cite{Peimbert1967,Peimbert1969}, for a given ion $X^{i+}$, the average temperature of the gas will be:
\begin{equation}
    \label{eq:topeimbert}
    T_0(X^{i+})=\frac{\int T_{\rm e} n_{\rm e} n(X^{i+}) dV }{\int n_{\rm e} n(X^{i+}) dV},
\end{equation}
where $T_{\rm e}$, $n_{\rm e}$ are the electron temperature and density at a given volume of the nebula, whereas $n(X^{i+})$ is the particle density of the ion. If there are temperature inhomogeneities, these can be quantified by the root mean square deviation from the averaged temperature:
\begin{equation}
    \label{eq:t2peimbert}
    t^2(X^{i+})=\frac{\int [T_{\rm e}-T_0(X^{i+})]^2 n_{\rm e} n(X^{i+}) dV }{T_0(X^{i+})^2\int n_{\rm e} n(X^{i+}) dV}.
\end{equation}

We use a collection of 20 and 23 deep optical spectra of Galactic and extragalactic H\,II regions, respectively as well as 8 spectra of Galactic ring nebulae (RNe), which are nebulae created by mass-loss episodes from young very massive stars. These data have the best quality available in the literature, being carefully and homogeneously reduced and analyzed over more than 20 years by our research group, containing the following characteristics: (i) have at least one line detected from the recombination O\,II multiplet V1, free of line blending with other lines that are not O\,II V1 RLs or observational issues. (ii) uncertainties in the line ratios smaller than 40\%. (iii) reliable detections of both temperature-sensitive [N\,II] $\lambda 5755$ and [O\,III] $\lambda 4363$ CELs. Following the so-called direct method, we calculate the electron density of the nebulae, as well as $T_{\rm e}$([O\,III] $\lambda 4363/\lambda 5007$), $T_{\rm e}$([N\,II] $\lambda 5755/\lambda 6584$) and $T_{\rm e}$([Ar\,III] $\lambda 5192/\lambda 7751$) when available. We derive $T_{\rm e}$(O\,II V1/[O\,III]$\lambda 5007$) using the formalism proposed by \cite{Peimbert2013} with the updated atomic data from \cite{Storey2017}. 

Not surprisingly, in all cases, $\text{O}^{2+}/\text{H}^{+}$ derived from the O\,II V1 RLs is found to be higher than the same ratio derived from the [O\,III] $\lambda 5007$ CEL (see Table~\ref{tab:ionicabundances}). The ratio of both ionic abundances is called the abundance discrepancy factor (ADF). By comparing the $T_{\rm e}$([O\,III] $\lambda 4363/\lambda 5007$) and $T_{\rm e}$(O\,II V1/[O\,III]$\lambda 5007$) -determined using Eqs. 10 and 11 from \cite{Peimbert2013}-, we derive $T_0(\text{O}^{2+})$ and $t^2(\text{O}^{2+})$. The last parameter is the degree of spatial temperature variations within the high ionization zone that is required in order to reconcile the $\text{O}^{2+}/\text{H}^{+}$ ratios derived from RLs and CELs.

\begin{figure}[h]
\centering
\includegraphics[width=\textwidth]{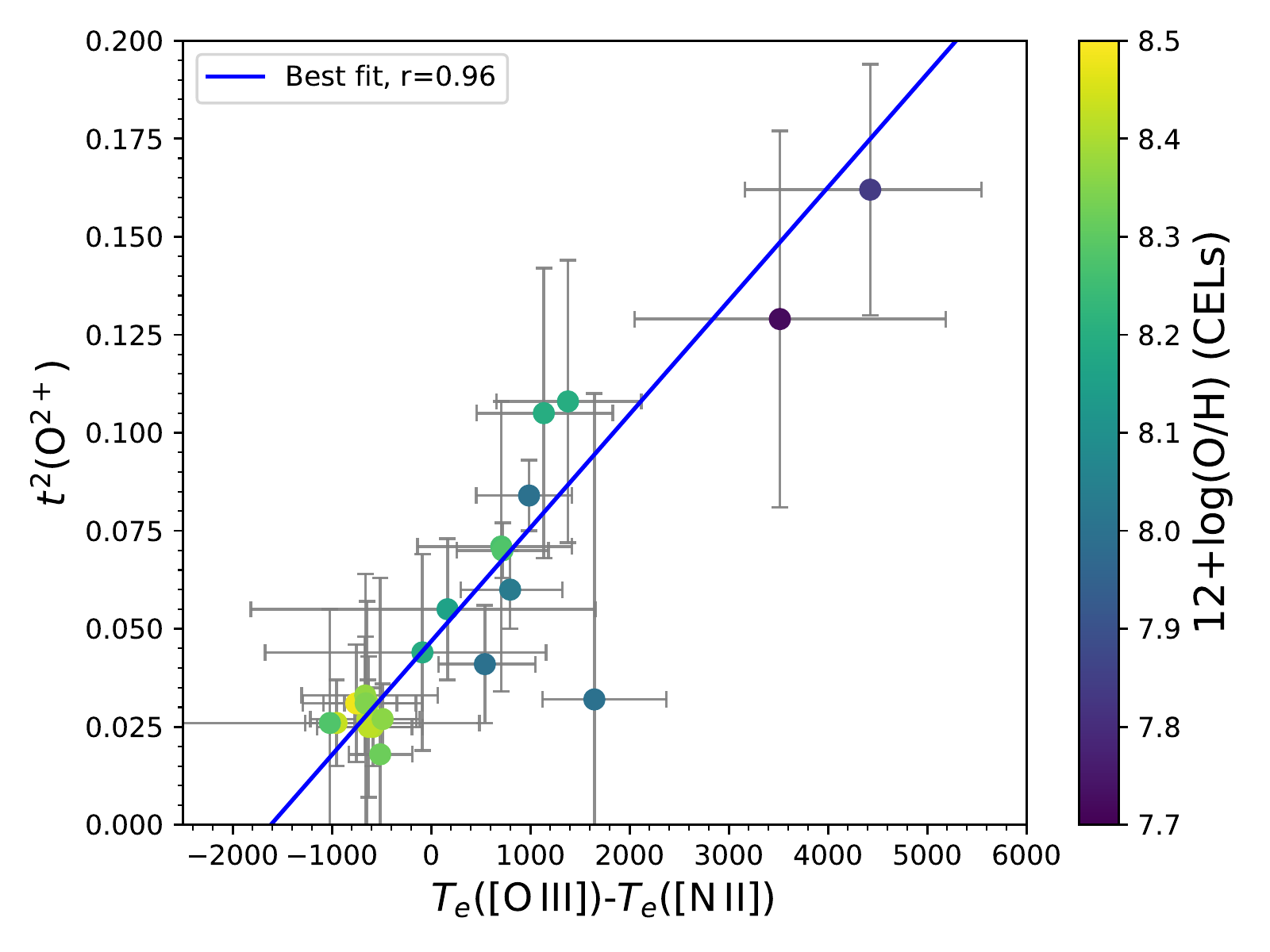}
\caption{Linear correlation between $t^2(\text{O}^{2+})$, the temperature inhomogeneities parameter value required to solve the abundance discrepancy problem and the difference of the derived temperatures of the ionized gas for extragalactic H\,II regions. The color bar corresponds to the O/H abundance derived from CELs assuming a homogeneous temperature structure ($t^2=0$). The Pearson correlation coefficient of the fit (r) is 0.96. }\label{Fig:Delta_t2}
\end{figure}

Since the observations of the extragalactic H\,II regions cover the complete temperature structure, we will initially focus our analysis on these regions in order to avoid the scatter that may be induced by aperture effects. Contrary to what was argued in several studies \cite{garciarojas2007,Peimbert2017,mendezdelgado22a} that were limited to a small number of H\,II regions, there is no ``typical'' $t^2(\text{O}^{2+})$ value for these nebulae (See Table~\ref{tab:thingsofOII}). As pointed out by \cite{esteban02,Peimbert2012,Toribio17}, very high values of $t^2(\text{O}^{2+})$ can be found in ``low metallicity'' H\,II regions. However, the connection between $t^2(\text{O}^{2+})$ and the physical conditions of the gas was not obvious in these previous studies. In Fig.~\ref{Fig:Delta_t2} we show the missing connection, $t^2(\text{O}^{2+})$ is linearly correlated with $\Delta T_{\rm e}=T_{\rm e}$([O\,III] $\lambda 4363/\lambda 5007)-T_{\rm e}$([N\,II] $\lambda 5755/\lambda 6584$), as follows:
\begin{equation}
    \label{eq:corre}
    t^2\left(\text{O}^{2+}\right)= \left( 2.90 \pm 0.21 \right) \times 10^{-5} \Delta T_{\rm e} + \left( 4.68 \pm 0.18 \right) \times 10^{-2} (\text{K}).  
\end{equation}

\begin{figure}[h]
\centering
\includegraphics[width=\textwidth]{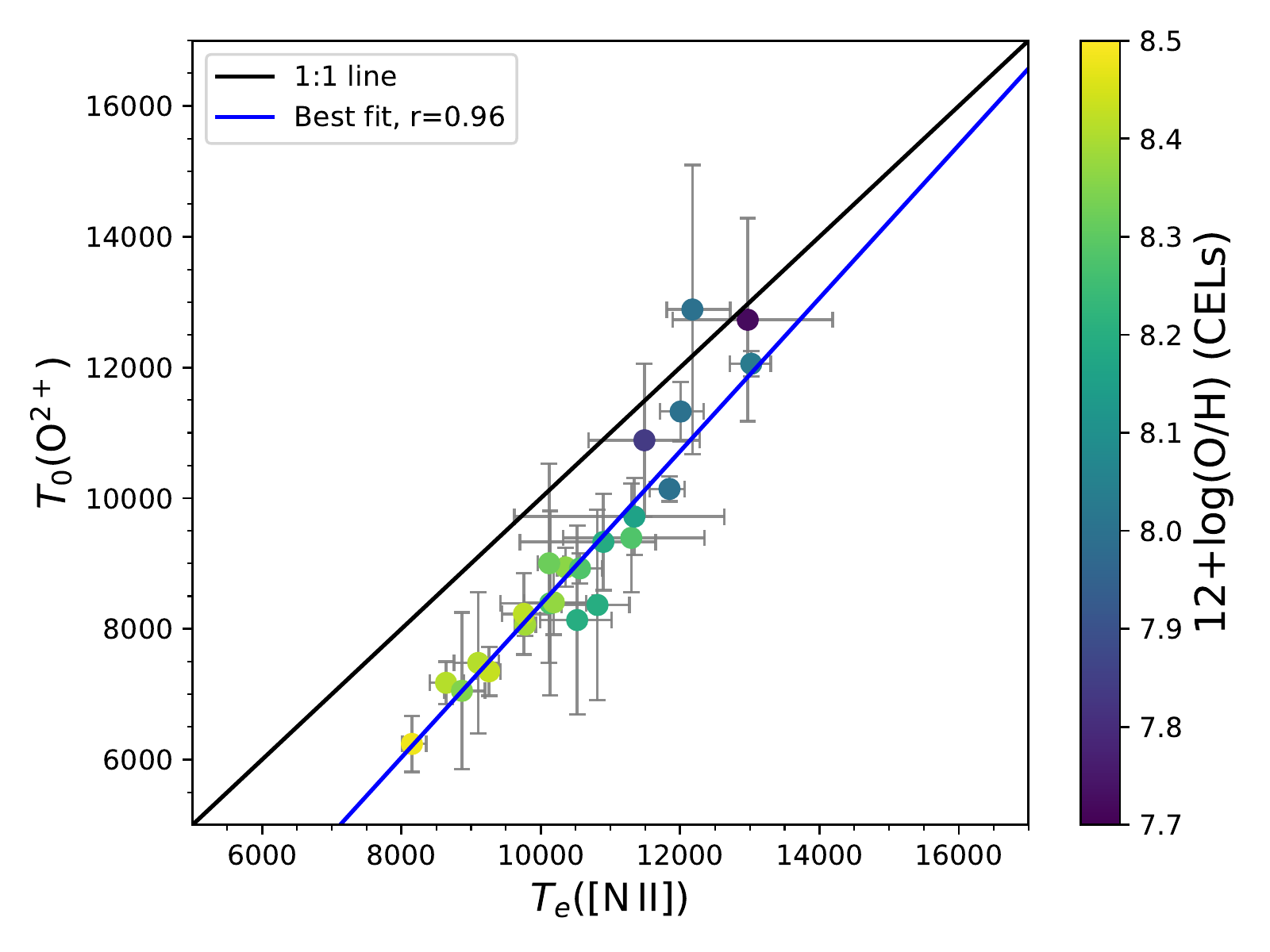}
\caption{Relation between $T_{\rm e}$([N\,II] $\lambda 5755/\lambda 6584$) and $T_0(\text{O}^{2+})$ for extragalactic H\,II regions.}\label{Fig:TN2_To}
\end{figure}

The explanation and the implications of this correlation are the following:

\begin{itemize}

    \item The linear relationship between $t^2(\text{O}^{2+})$ and $\Delta T_{\rm e}$ is a consequence of the fact that $T_{\rm e}$([N\,II] $\lambda 5755/\lambda 6584$) behaves as the average temperature in the N$^+$ volume (See Eq.~\eqref{eq:t2peimbert}). This implies that temperature inhomogeneities are negligible in the low ionization zone (i.e. $t^2(\text{O}^{+}) \approx 0$). This is demonstrated in Fig.~\ref{Fig:TN2_To}, where a tight linear correlation between $T_{\rm e}$([N\,II] $\lambda 5755/\lambda 6584$) and $T_0(\text{O}^{2+})$ is presented.

    \item The fact that $T_{\rm e}$(O\,II V1/[O\,III]$\lambda 5007$) (practically identical to $T_0(\text{O}^{2+})$ in H\,II regions) is correlated with $T_{\rm e}$([N\,II] $\lambda 5755/\lambda 6584$) also demonstrates that the first value is physically representing a gas temperature.

    \item As $T_{\rm e}$(O\,II V1/[O\,III]$\lambda 5007$) $<$ $T_{\rm e}$([O\,III] $\lambda 4363/\lambda 5007)$, with the first being a real value of temperature, we show that a physical process is acting preferentially in the inner high ionization volume, increasing the intensity of the temperature sensitive [O\,III] $\lambda 4363$ CEL. This demonstrate the existence of real temperature inhomogenities.

    \item Therefore $t^2(\text{O}^{2+})$ is not an ad-hoc parameter to solve the abundance discrepancy of the O$^{2+}$ ion, it is indeed describing inhomogenities in the temperature of the highly ionized gas. 

    
\end{itemize}

On the first point mentioned above, it should be noted that $T_{\rm e}$([N\,II] $\lambda 5755/\lambda 6584$) is not involved in the estimation of $T_0(\text{O}^{2+})$ or $T_{\rm e}$(O\,II V1/[O\,III]$\lambda 5007$) in any sense.  We interpret the relationship shown in Fig.~\ref{Fig:TN2_To} as a consequence of the natural stratification of temperatures expected in the photoionized gas due to the combination of the hardening of the stellar radiation and the stronger cooling from [O\,III] lines, which are more efficient at cooling than those of [O\,II] \citep{stasinska:1980,Garnett:1992}. The linear fit between both quantities is presented in Eq.~\eqref{eq:To_TN_fit}.

\begin{equation}
    \label{eq:To_TN_fit}
    T_0(\text{O}^{2+})=\left( 1.17 \pm 0.05 \right) \times T_{\rm e}(\text{[N\,II]}) - \left( 3340 \pm 470 \right) (\text{K}).  
\end{equation}

\begin{figure}[h]
\centering
\includegraphics[width=\textwidth]{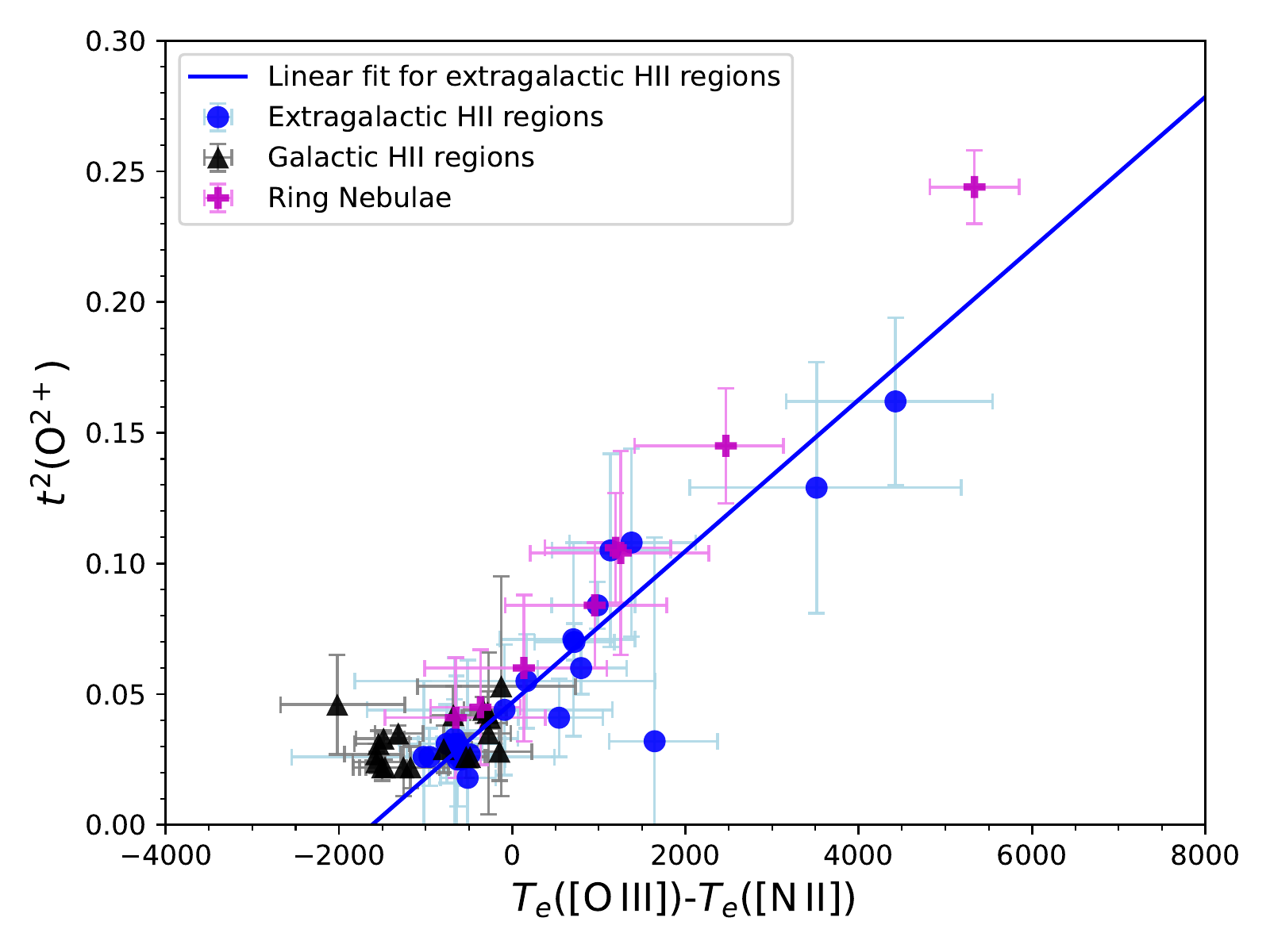}
\caption{$t^2(\text{O}^{2+})$ versus $\Delta T_{\rm e}$ relation for Galactic and extragalactic H\,II regions and ring nebulae.}\label{Fig:Delta_t2-2}
\end{figure} 

When we extend the analysis to the Galactic H\,II regions and RNe, as shown in Fig.~\ref{Fig:Delta_t2-2}, we observe that in general they follow the same trend as the extragalactic H\,II regions. Five spectra from different areas of NGC\,6888 \cite{esteban16}, a bubble of ionized gas created by the Wolf-Rayet WN6 star WR136 and three from NGC\,7635 \cite{esteban16}, created by the O6.5(n)fp star BD+60 2522 \cite{sota11} and the surrounding interstellar gas show also a good agreement with the general trend of the H\,II regions. This is remarkable in the case of some peripheral zones of NGC\,6888, that show particularly large values of $t^2(\text{O}^{2+})$. This implies that stellar feedback processes can contribute to the inhomogeneous heating of the gas and are related to the origin of temperature inhomogeneities. This is consistent with the fact that the gas with a high degree of ionization, which is located physically closer to the ionizing stars in typical H\,II regions, is the most affected one by the temperature inhomogeneities. Variations in the hardness of the stellar ionizing spectra may also contribute to $t^2(\text{O}^{2+})$ \cite{Perez1997,Ercolano07}.

\begin{figure}[h]
\centering
\includegraphics[width=\textwidth]{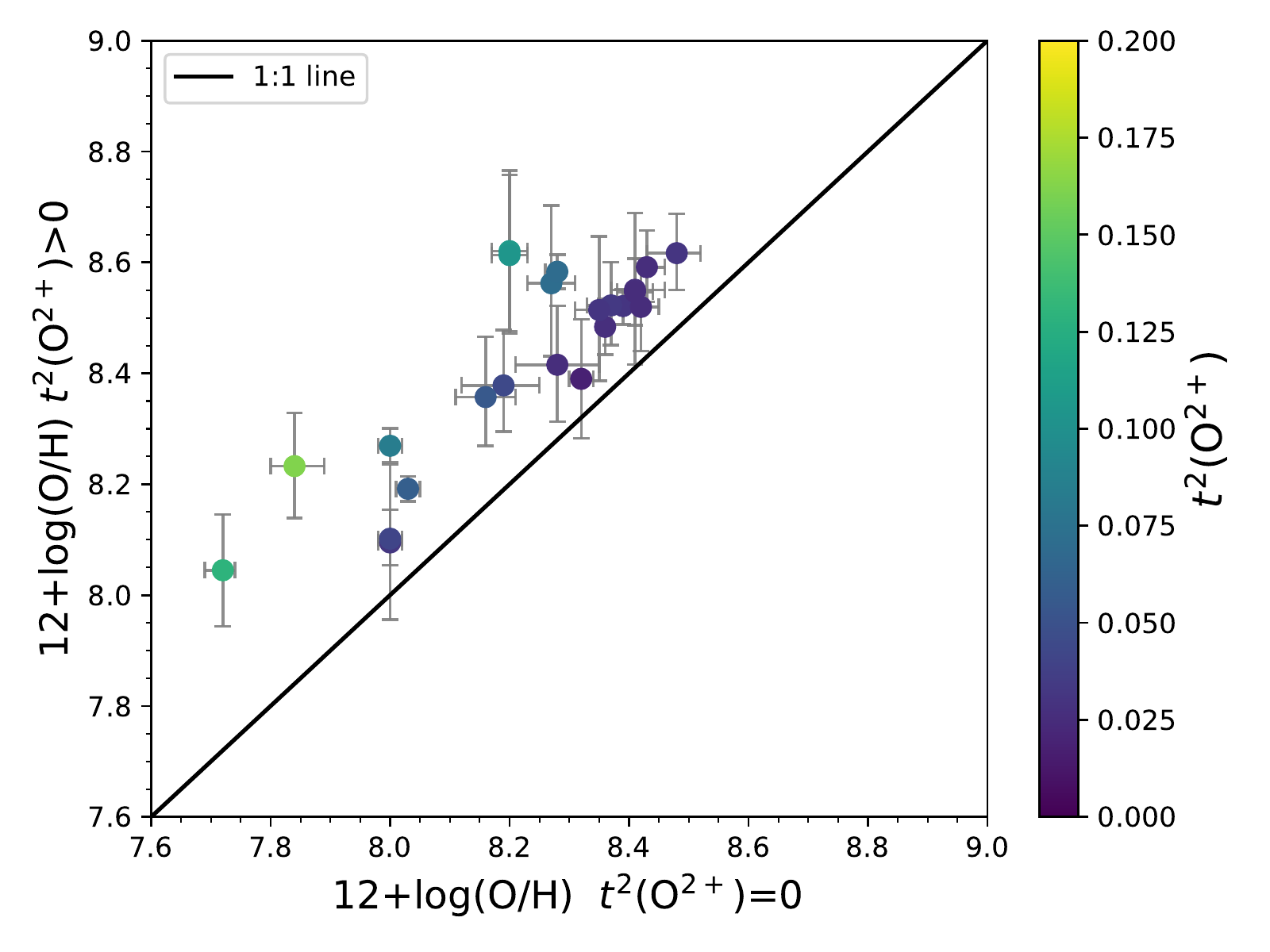}
\caption{Effects on metallicity of considering temperature inhomogeneities affecting only the high ionization volume in our sample of extragalactic H\,II regions.}\label{Fig:impact}
\end{figure}

\begin{figure}[h]
\centering
\includegraphics[width=\textwidth]{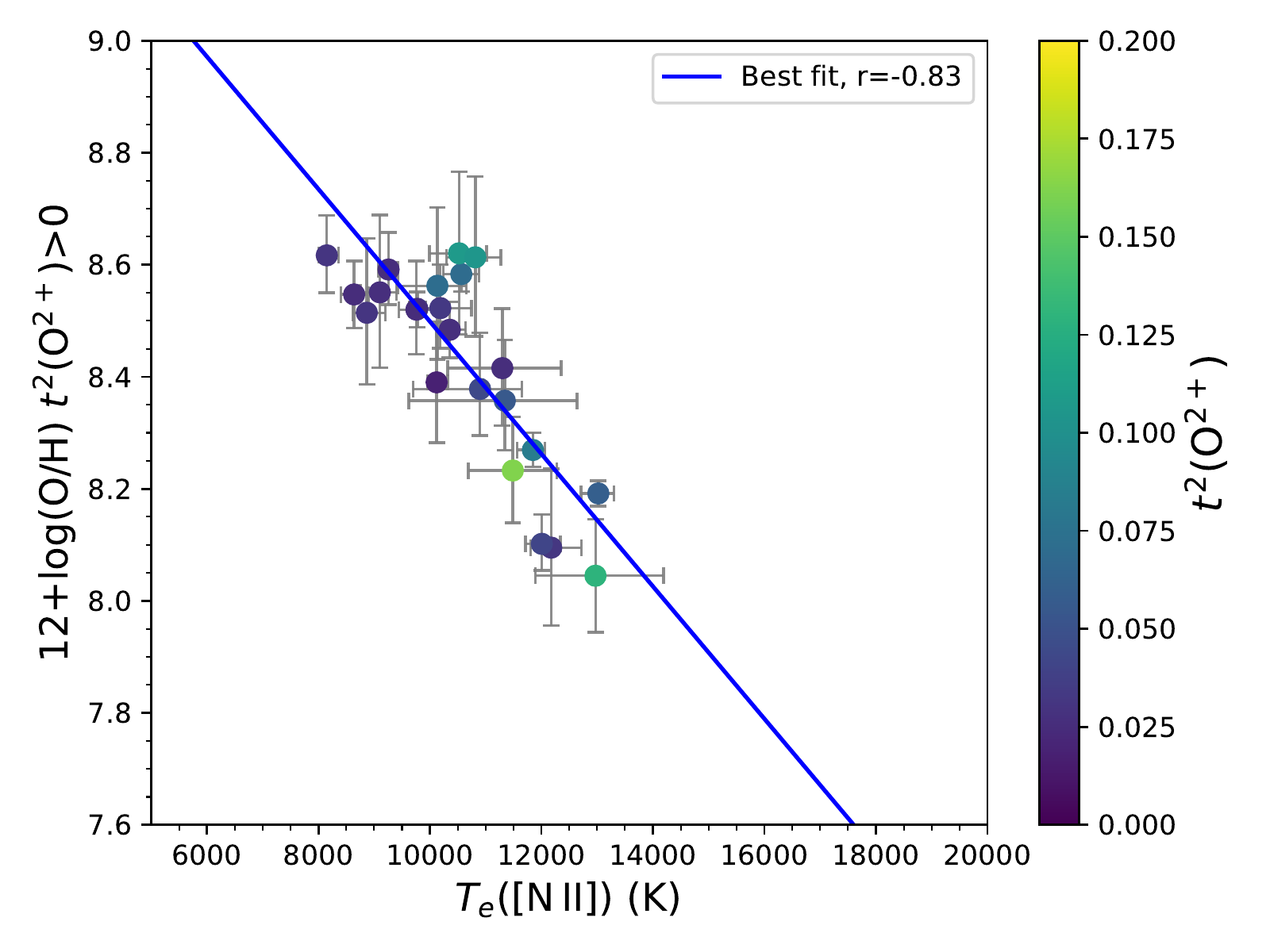}
\caption{$T_{\rm e}$([N\,II] $\lambda 5755/\lambda 6584$)-metallicity relation considering $t^2(\text{O}^{2+})>0$ for our sample of extragalactic H\,II regions.}\label{Fig:temN2-metal}
\end{figure}

To test the impact of our results on the metallicity of our sample of extragalactic H\,II regions, we compare the effect of considering $t^2>0$ only in the high ionization volume. The results are shown in Fig.~\ref{Fig:impact}, and highlight that the metallicity underestimation tends to be larger at lower metallicities, reaching values as high as 0.5 dex. Considering these results, we can infer a $T_{\rm e}$([N\,II] $\lambda 5755/\lambda 6584$)-metallicity relation, provided in Eq.~\eqref{eq:metal-tem} and shown in Fig.~\ref{Fig:temN2-metal}, which allows a quick estimation of the metallicity considering $t^2(\text{O}^{2+})>0$.

\begin{equation}
    \label{eq:metal-tem}
    12+\text{log(O/H)}=\left(-1.19 \pm 0.14 \right)\times 10^{-4} T_{\rm e} \text{([N\,II])} + \left( 9.68 \pm 0.15\right) \text{(K).}
\end{equation}

Based on deep optical spectra of Galactic and extragalactic H\,II regions and ring nebulae, we conclude that temperature inhomogenities are present within these nebulae. These inhomogeneities cause the long-standing discrepancy between the heavy-element abundances determined with CELs and RLs. The observational evidence show that the temperature inhomogenities affect only the volume of high degree of ionization, the closest to the ionizing stars. This is a change of paradigm even for the community who consider $t^2>0$ in their studies. Stellar feedback processes and radiation hardness fluctuations of the stellar spectra are potential candidates to produce the observed inhomogeneities. The chemical abundances estimated with CELs, the most accessible observationally, must be revised because they are underestimated. Both the abundances determined by the direct method and those estimated by strong lines ones calibrated by observations require revision, since both depend on $T_{\rm e}$([O\,III] $\lambda 4363/\lambda 5007$). We provide Eq.~\eqref{eq:corre}, Eq.~\eqref{eq:To_TN_fit} and Eq.~\eqref{eq:metal-tem}, which allow us to consider the effect of temperature inhomogenities in the calculation of chemical abundances even if only CELs are observed. This is especially important in low metallicity H\,II regions where the values of $t^2(\text{O}^{2+})$ are higher, and metallicities are systematically underestimated (Fig.~\ref{Fig:impact}). This work has far-reaching implications for chemical evolution studies of galaxies, where the build-up of elements is treated as a cosmic clock.  This is particular critical in our examination of the mass-metallicity relationships of galaxies \cite{Maiolino19}, as well as recent efforts to directly track chemical evolution over cosmic time by using JWST to observe [O\,III] auroral lines in high-z galaxies \cite{arellano2022,Curti2023,Katz:2023}.  Based on our findings, $t^2$ should be considered as a fundamental physical parameter both in observational and theoretical works. 



\section{Methods}\label{methods}

For the sample of nebulae presented in Tables~\ref{tab:physical_cond_sampleextra}, \ref{tab:physical_cond_sampleRNe}, \ref{tab:densities_sampleGal} and \ref{tab:densities_samplePNe}, we derive the physical conditions and ionic abundances by the so-called direct method in a consistent way. All values obtained are based on the intensities reported in the reference spectra. We discard the use of lines with observational issues as telluric absorptions or sky line contaminations. We use PyNeb 1.1.13 \citep{Luridiana:2015} and the atomic data set provided in Table~\ref{tab:atomic_data} to estimate the electron density ($n_{\rm e}$) and temperature ($T_{\rm e}$). As a first step, we derive the $n_{\rm e}$-$T_{\rm e}$ cross-match convergence by solving the statistical equilibrium equations with the [S\,II] $\lambda 6731/\lambda 6716$, [O\,II] $\lambda 3726/\lambda 3729$, [Cl\,III] $\lambda 5538/ \lambda 5518$, [Fe\,III] $\lambda 4658/ \lambda 4702$ and [Ar\,IV] $\lambda 4740/ \lambda 4711$ density diagnostics and the [O\,III] $\lambda 4363/ \lambda 5007$, [N\,II] $\lambda 5755/ \lambda 6584$, [Ar\,III] $\lambda 5192/ \lambda 7135$ temperature diagnostics, using the task {\it getCrossTemDen} of PyNeb. For each density diagnostic, we average the convergence value of each cross-matching. Considering that [S\,II] $\lambda 6731/\lambda 6716$ and [O\,II] $\lambda 3726/\lambda 3729$ have their maximum sensitivity with density in the range of values $10^2 \text{ cm}^{-3}<n_{\rm e}<10^4 \text{ cm}^{-3}$ \cite{IFF05,Tayal:2010,FroeseFischer:2004,Kisielius:2009}, whereas [Cl\,III] $\lambda 5538/ \lambda 5518$, [Fe\,III] $\lambda 4658/ \lambda 4702$ and [Ar\,IV] $\lambda 4740/ \lambda 4711$ are good diagnostics in the range of $10^3 \text{ cm}^{-3}<n_{\rm e}<10^6 \text{ cm}^{-3}$ \cite{Fritzsche:1999,Butler:1989,Mendoza:1982b,Ramsbottom:1997,Quinet:1996,Zhang:1996}, we derive an average representative $n_{\rm e}$ adopting the following criteria:

\begin{itemize}
    \item If $n_{\rm e}$([S\,II]) $<$ 100 cm$^{-3}$, we adopt $n_{\rm e}<$ 100 cm$^{-3}$.
    \item If 100 cm$^{-3}$ $<$ $n_{\rm e}$([S\,II]) $<$ 1000 cm$^{-3}$, we adopt the average value of $n_{\rm e}$([S\,II]) and $n_{\rm e}$([O\,II]).
    \item If $n_{\rm e}$([S\,II]) $>$ 1000 cm$^{-3}$, we take the average values of $n_{\rm e}$([S\,II]), $n_{\rm e}$([O\,II]), $n_{\rm e}$([Cl\,III]), $n_{\rm e}$([Fe\,III]) and $n_{\rm e}$([Ar\,IV]) when available.
\end{itemize}

In all cases, the averaged values are weighted by the inverse of the square of the error. All the extragalactic H\,II regions but one follow the first two criteria and therefore in Tables~\ref{tab:physical_cond_sampleextra}, \ref{tab:physical_cond_sampleextra_others} we only present $n_{\rm e}$([O\,II]), $n_{\rm e}$([S\,II]) as well as the adopted value. The exception is N\,88A \cite{dominguezguzman22}, where we follow the third criteria given its higher density. Therefore, in addition to the values of $n_{\rm e}$([S\,II]) and $n_{\rm e}$([O\,II]) shown in Table~\ref{tab:physical_cond_sampleextra}, we take into account $n_{\rm e}$([Cl\,III])=$3890^{+280} _{-320}$ cm$^{-3}$, $n_{\rm e}$([Ar\,IV])=$5860^{+910} _{-920}$ cm$^{-3}$ and $n_{\rm e}$([Fe\,III])=$7570^{+3870} _{-2940}$ cm$^{-3}$ for N\,88A. Once $n_{\rm e}$ is well established, we derive $T_{\rm e}$([O\,III] $\lambda 4363/\lambda 5007$), $T_{\rm e}$([N\,II] $\lambda 5755/\lambda 6584$) and $T_{\rm e}$([Ar\,III] $\lambda 5192/\lambda 7751$) when available, using the {\it getTemDen} task of PyNeb.

In the case of O\,II RLs, we consider only those from the multiplet V1, individually measured or blended with other lines of the same multiplet. In each region we sum the intensity of all the O\,II V1 observed lines and we estimate the intensity of the complete multiplet with the effective recombination coefficients from \cite{Storey2017}, which consider the population of the O\,II levels. These values of the total intensity of multiplet V1 are presented in the fifth column of the Table~\ref{tab:thingsofOII}. The number of observed O\,II V1 lines is presented in the sixth column. Note that the relative intensity of the O\,II V1 RLs is independent of the temperature, while it depends on the density for values between $10^{2.5} \text{ cm}^{-3}<n_{\rm e}<10^4 \text{ cm}^{-3}$ \cite{Storey2017}. Since the total intensity of the multiplet V1 is independent of the density, this could introduce some errors in high-density regions with an incorrect estimation of $n_{\rm e}$ (by an order of magnitude) and when only one O\,II V1 RL is detected. However, this case does not occur in our sample and our conclusions do not depend on this.

$T_{\rm e}$(O\,II V1/[O\,III]$\lambda 5007$) was derived using the aforementioned O\,II V1 intensities with respect to [O\,III]$\lambda 5007$ and is presented in the second column of Table~\ref{tab:thingsofOII} for each object of our sample. The O\,II V1/[O\,III]$\lambda 5007$ line intensity ratio depends on the electron temperature in a different proportion than  [O\,III] $\lambda 4363$/$\lambda 5007$, given the different dependence of the emissivities of O\,II V1 and [O\,III] $\lambda 4363$ lines with $T_{\rm e}$. If we assume that both RLs and CELs arise from the same ionized gas, the difference between $T_{\rm e}$(O\,II V1/[O\,III]$\lambda 5007$) and $T_{\rm e}$([O\,III] $\lambda 4363/\lambda 5007$) give us an estimate of the temperature inhomogeneities. It should be noted that the derived $T_{\rm e}$(O\,II V1/[O\,III]$\lambda 5007$) is the temperature value that eliminates the discrepancy between the $n(\text{O}^{2+})$/$n(\text{H}^{+})$ ratio derived with CELs and RLs. Using the formalism of \cite{Peimbert2013} -Eqs. (10) and (11)-, we calculate $T_0$(O$^{2+}$) and $t^2$(O$^{2+}$). It is important to highlight that the ADF is not synonymous with $t^2$. ADF(O$^{2+}$) is the consequence of an inhomogeneous temperature structure affecting $T_{\rm e}$([O\,III] $\lambda 4363/\lambda 5007$) and therefore both quantities are intrinsically related and described by both $T_0$(O$^{2+}$) and $t^2$(O$^{2+}$) \cite{Peimbert1967}.

\begin{figure}[h]
\centering
\includegraphics[width=\textwidth]{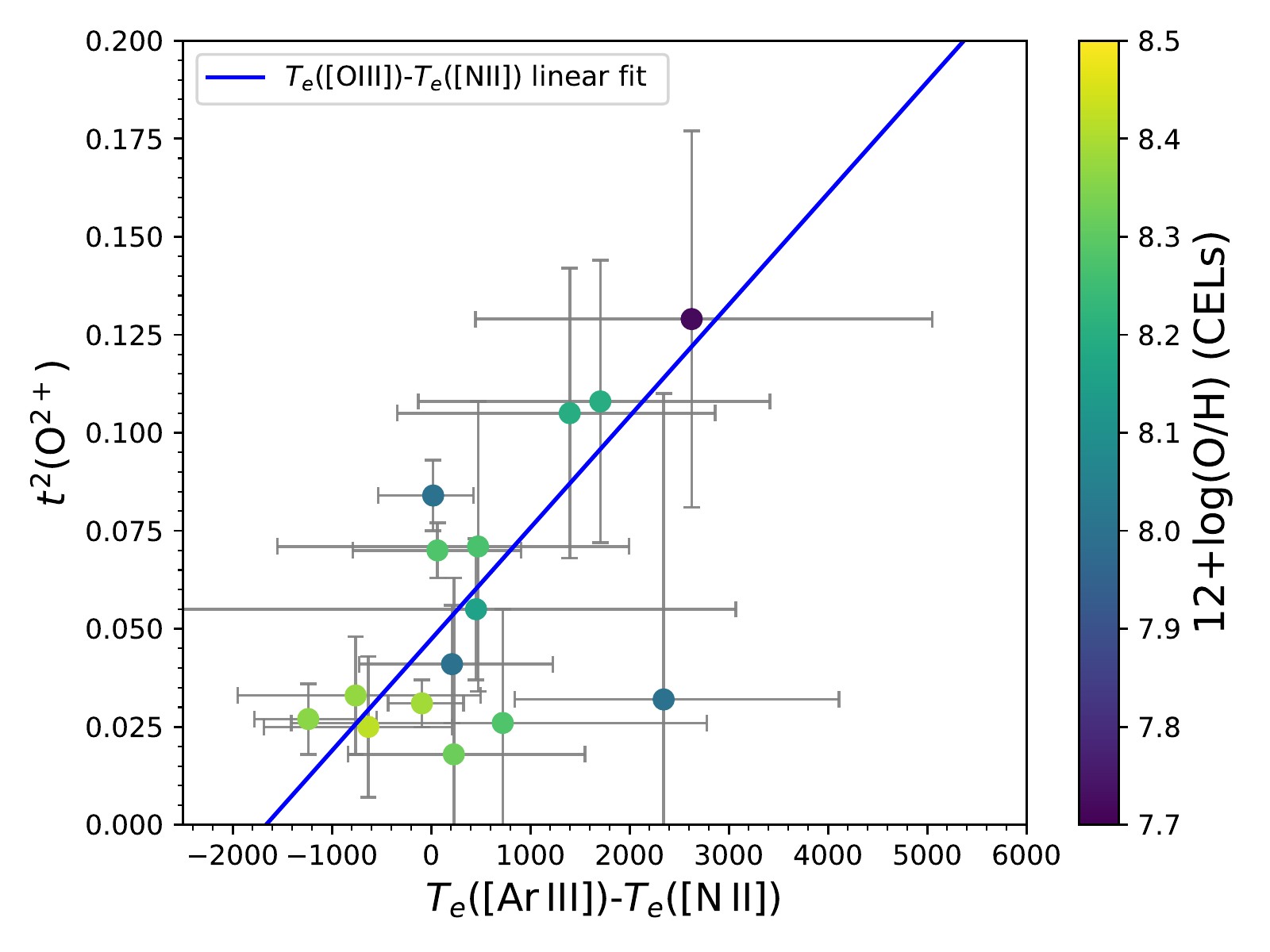}
\caption{$t^2(\text{O}^{2+})$ compared with $T_{\rm e}$([Ar\,III] $\lambda 5192/\lambda 7751)-T_{\rm e}$([N\,II] $\lambda 5755/\lambda 6584$) for the studied sample of extragalactic H\,II regions. }\label{Fig:Delta_t2-Ar3}
\end{figure}

In the first part of this article, we show a tight relationship between $t^2(\text{O}^{2+})$ and $\Delta T_{\rm e}=T_{\rm e}$([O\,III] $\lambda 4363/\lambda 5007)-T_{\rm e}$([N\,II] $\lambda 5755/\lambda 6584$). As mentioned, this is a consequence of the fact that $T_{\rm e}$([N\,II] $\lambda 5755/\lambda 6584$) is linearly correlated with $T_0$(O$^{2+}$) (or $T_{\rm e}$(O\,II V1/[O\,III]$\lambda 5007$) as they are practically equivalent in the analyzed H\,II regions). Therefore, Fig.~\ref{Fig:Delta_t2} and Fig.~\ref{Fig:TN2_To} are complementary. Mathematically, the formalism of \cite{Peimbert2013} predicts the linear fit shown in Eq.~\eqref{eq:corre} with the presence of any phenomenon that produces $T_{\rm e}$([O\,III] $\lambda 4363/\lambda 5007$) $>$ $T_{\rm e}$(O\,II V1/[O\,III]$\lambda 5007$) (and consequently ADF(O$^{2+}$)$>$0) and preserves the linearity between $T_0$(O$^{2+}$) and $T_{\rm e}$([N\,II] $\lambda 5755/\lambda 6584$). Temperature inhomogeneities in the volume of high degree of ionization is the only paradigm that fulfill these requirements as other explanations to the ADF(O$^{2+}$) would imply the invalidity of $T_{\rm e}$(O\,II V1/[O\,III]$\lambda 5007$) as a real temperature diagnostic. We discard other explanations to the ADF(O$^{2+}$) in H\,II regions from the literature as follows:

\begin{itemize}

    \item If there were chemical inhomogeneities, they would create temperature variations. Nevertheless, in that case, $T_{\rm e}$(O\,II V1/[O\,III]$\lambda 5007$) would not correlate with $T_{\rm e}$([N\,II] $\lambda 5755/\lambda 6584$) as the volumes emitting RLs and CELs would be different \cite{Zhang:2007}.
    
   \item If the ADF(O$^{2+}$) were created by fluorescent excitation of the O\,II V1 RLs, then these lines would be enhanced in proportion to parameters different than the electron temperature of the gas (such as the effective temperature of the ionizing star or the optical depth). Therefore $T_{\rm e}$(O\,II V1/[O\,III]$\lambda 5007$) and $T_{\rm e}$([N\,II] $\lambda 5755/\lambda 6584$) would not correlate.

   \item If the free electrons follow a kappa distribution instead of a Maxwellian one, supra-thermal electrons would create temperature variations that can be directly described by $t^2$. However, both observational and theoretical evidence indicate the prevalence of a Maxwellian distribution of velocities \cite{Ferland2016}.
    
   \item In the case of errors in the recombination coefficients of the O\,II V1 RLs, the observed ADF(O$^{2+}$) would be approximately constant\cite{rodriguez2010}, which is not the case. 

   \item In order to have an important recombination contribution to the auroral [O\,III] $\lambda 4363$ line intensity, a large fraction of O$^{3+}$/O is required. This is not the case in the analyzed H\,II regions. Moreover, Fig.~\ref{Fig:Delta_t2-Ar3} shows that $t^2(\text{O}^{2+})$ and $T_{\rm e}$([Ar\,III] $\lambda 5192/\lambda 7751)-T_{\rm e}$([N\,II] $\lambda 5755/\lambda 6584$) also have a correlation despite of the larger scatter. 

    \end{itemize}


\begin{figure}[h]
\centering
\includegraphics[width=\textwidth]{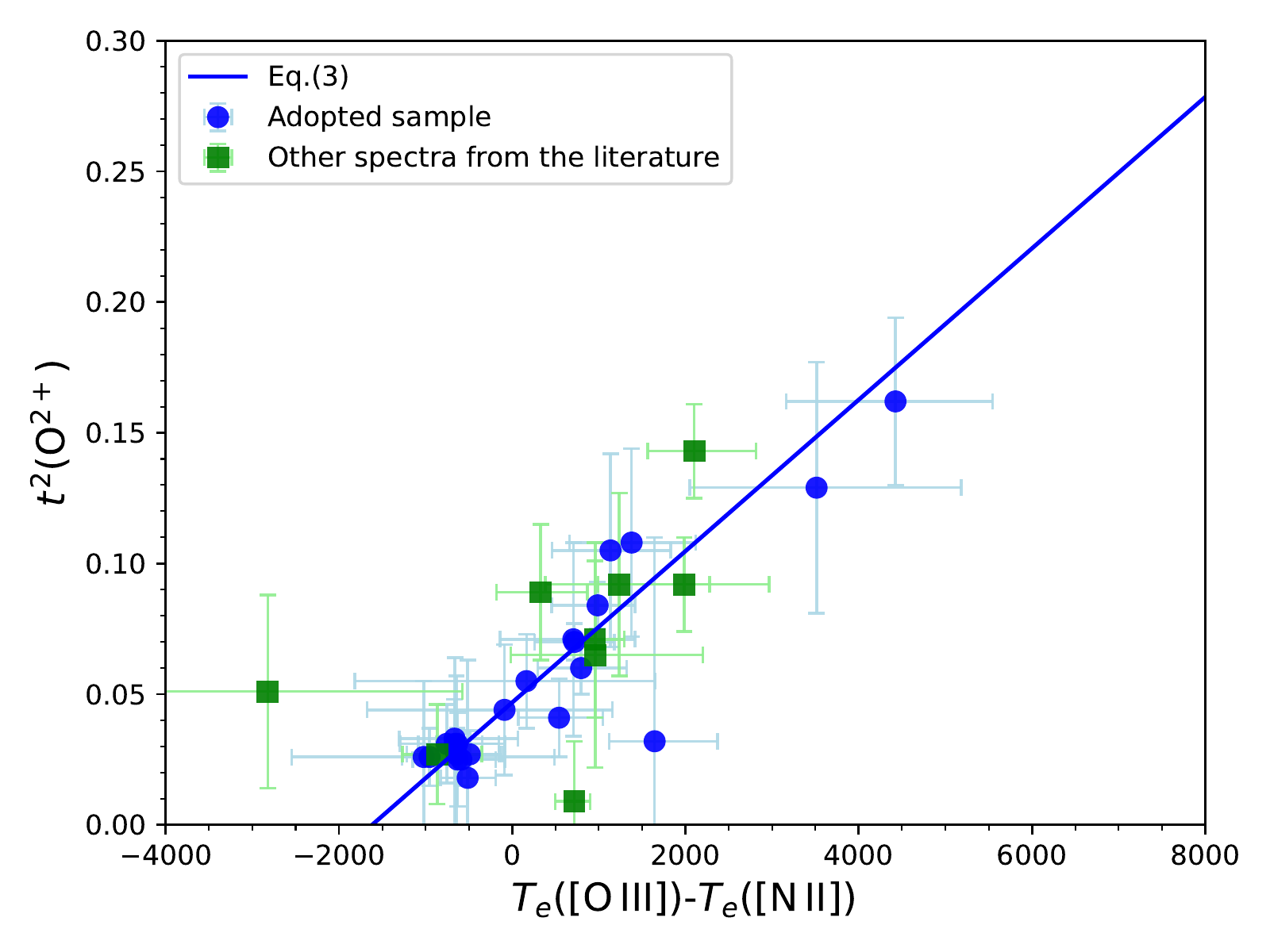}
\caption{$t^2(\text{O}^{2+})$ versus $\Delta T_{\rm e}$ for the extragalactic H\,II regions of our sample (blue dots) and those from the literature presented in Table~\ref{tab:physical_cond_sampleextra_others} (green squares).}\label{Fig:Delta_t2others}
\end{figure}

To ensure that our conclusions are free of selection bias, we searched for spectra of extragalactic H\,II regions in the literature that follow the same selection criteria of our study. Due to the difficulty of simultaneously observing [N\,II] $\lambda 5755$, [O\,III] $\lambda 4363$ and the O\,II V1 RLs, there are not many candidates. We study the 9 spectra presented in Table~\ref{tab:physical_cond_sampleextra_others} by analyzing them with the same methodology as for the rest of the nebulae. In Fig.~\ref{Fig:Delta_t2others} we show that our conclusions do not depend on the adopted sample as the H\,II regions from the literature follow the same trend.

\begin{figure}[h]
\centering
\includegraphics[width=\textwidth]{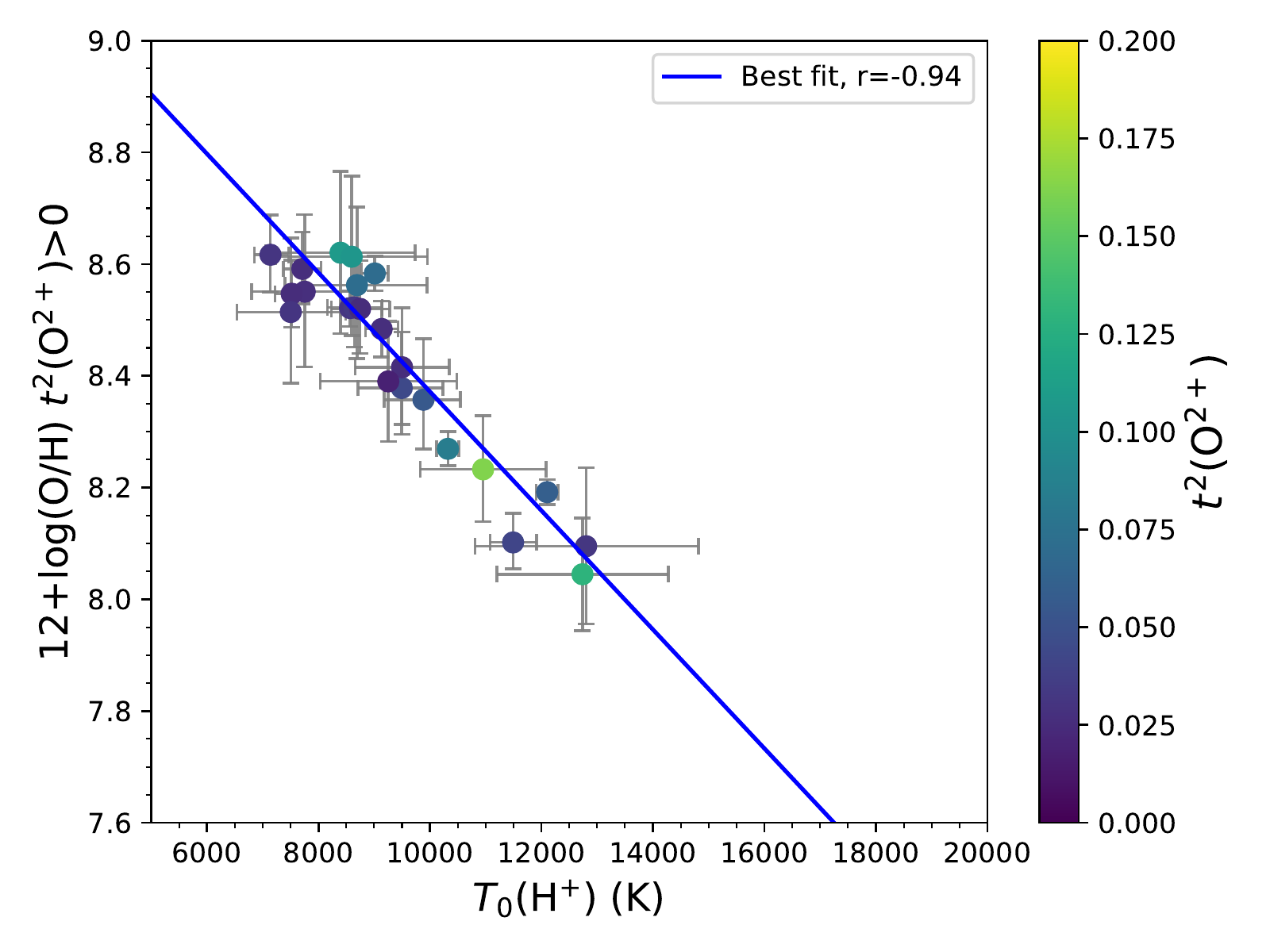}
\caption{$T_{\rm 0}(\text{H}^{+})$-metallicity relation considering $t^2(\text{O}^{2+})>0$ for our sample of extragalactic H\,II regions.}\label{Fig:temN2-metalHI}
\end{figure}

In Table~\ref{tab:ionicabundances} we present the O$^{+}$/H$^{+}$ (CELs), O$^{2+}$/H$^{+}$(CELs) and O$^{2+}$/H$^{+}$(RLs) ratios. The ionic abundances are derived with the {\it getIonAbundance} task of PyNeb. In the case of O$^{+}$/H$^{+}$, we use the [O\,II] $\lambda 3726+29$ intensity and $T_{\rm e}$([N\,II] $\lambda 5755/\lambda 6584$). To derive O$^{2+}$/H$^{+}$(CELs), we use [O\,III] $\lambda 5007$ and $T_{\rm e}$([O\,III] $\lambda 4363/\lambda 5007$), whereas to derive O$^{2+}$/H$^{+}$(RLs) we use the O\,II V1 intensity and $T_{\rm e}$([O\,III] $\lambda 4363/\lambda 5007$) (the temperature dependence is negligible). 
As H\,I RLs are emitted both in the high and low ionization volumes, we can apply the following relation \cite{Peimbert2000}:
\begin{equation}
    \label{eq:TH}
    T_{\rm 0}(\text{H}^{+})\approx\frac{T_{\rm 0}(\text{O}^{+})\times n(\text{O}^{+}) +T_{\rm 0}(\text{O}^{2+}) \times n(\text{O}^{2+})}{n(\text{O}^{+})+n(\text{O}^{2+})},
\end{equation}
where $T_{\rm 0}(\text{O}^{+})$ and $n(\text{O}^{+})$ are the average temperature and particle density of O$^{+}$, respectively. As our results show that $T_{\rm e}$([N\,II] $\lambda 5755/\lambda 6584$)$\approx T_{\rm 0}(\text{O}^{+})$, we can infer a $T_{\rm 0}(\text{H}^{+})$-metallicity relation, as is presented in Fig.~\ref{Fig:temN2-metalHI} and Eq.~\eqref{eq:metal-temPES}. This last relation is very important as it permits to estimate the metallicity from radio observations, where it is possible to measure $T_{\rm e}$(H\,I).

\begin{equation}
    \label{eq:metal-temPES}
    12+\text{log(O/H)}=\left(-1.07 \pm 0.09 \right)\times 10^{-4} T_{\rm 0} (\text{H}^{+}) + \left( 9.44 \pm 0.08\right) \text{(K).}
\end{equation}

\begin{figure}[h]
\centering
\includegraphics[width=\textwidth]{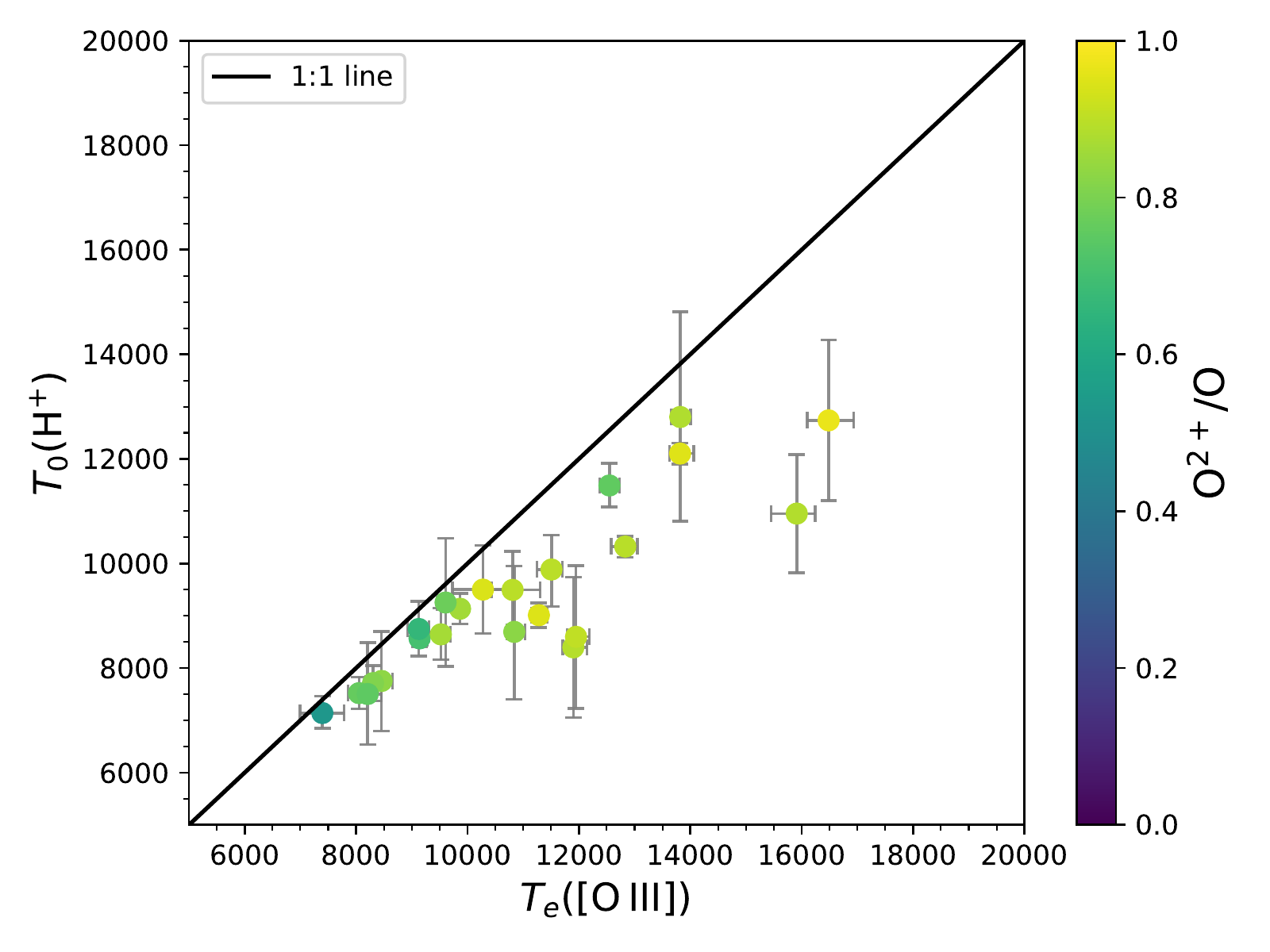}
\caption{Comparison between $T_{\rm e}$([O\,III] $\lambda 4363/\lambda 5007$) and $T_{\rm 0}(\text{H}^{+})$, which was derived using Eq.~\eqref{eq:TH} for our sample of extragalactic H\,II regions.}\label{Fig:temperaturasHI-O3}
\end{figure}

When $T_{\rm e}$([N\,II] $\lambda 5755/\lambda 6584$)$\approx T_{\rm 0}(\text{O}^{+})$ is considered, Eq.~\eqref{eq:TH} shows that it is possible to get $T_{\rm 0}(\text{H}^{+})$ values which are close to $T_{\rm e}$([O\,III] $\lambda 4363/\lambda 5007$) even if $t^2$(O$^{2+}$)$>0$. This is due to the fact that in the regions of low degree of ionization, the emission of H\,I RLs will come essentially from the volume of O$^{+}$, where temperature inhomogeneities are negligible. On the other hand, $T_{\rm e}$([N\,II] $\lambda 5755/\lambda 6584$)$\approx T_{\rm 0}(\text{O}^{+})$ have larger values than $T_{\rm 0}(\text{O}^{2+})$ (see Fig.~\eqref{Fig:TN2_To}). Therefore, finding some nebulae where $T_{\rm e}$(H\,I) $\approx T_{\rm e}$([O\,III]) is not an argument against the existence of temperature inhomogeneities. In Fig.~\ref{Fig:temperaturasHI-O3} we show that in some regions of our sample we could expect that $T_{\rm e}$(H\,I)$\approx$$T_{\rm e}$([O\,III] $\lambda 4363/\lambda 5007$) even if $t^2$(O$^{2+}$)$>0$. 

An important point to mention is the possible existence of ADF(O$^+$). To date, there are no reported ADF(O$^+$) (or from any ion of low degree of ionization) in extragalactic H\,II regions. The available cases are limited to Galactic planetary nebulae (PNe) and 3 Galactic H\,II regions: M8, M20 and the Orion Nebula \cite{garciarojas2007}. Besides the possible contribution of sky contamination in the O\,I 7773+ multiplet, it is necessary to study the aperture effects on these nebulae, since echelle observations \cite{esteban04,garciarojas06,garciarojas07} are limited to very small areas of the nebula. For instance, in the dense gas of the Orion Nebula, a high density clump (such as a proplyd or an Herbig-Haro object) can create a fake local peak of $T_{\rm e}$([N\,II] $\lambda 5755/\lambda 6584$) \cite{mesadelgado08,mendezdelgado21b}, giving the $n_{\rm e}$-sensitivity of [N\,II] $\lambda 5755/\lambda 6584$ at values higher than 1000 cm$^{-3}$. In any case, the ADF(O$^+$) could have a negligible effect on the integrated spectrum where the entire temperature structure is encompassed. If a large ADF(O$^+$) exists in extragalactic H\,II regions, this work shows that the main cause would not be temperature inhomogeneities. But before discussing a problem, we need evidence that it really exists.

\begin{figure}[h]
\centering
\includegraphics[width=\textwidth]{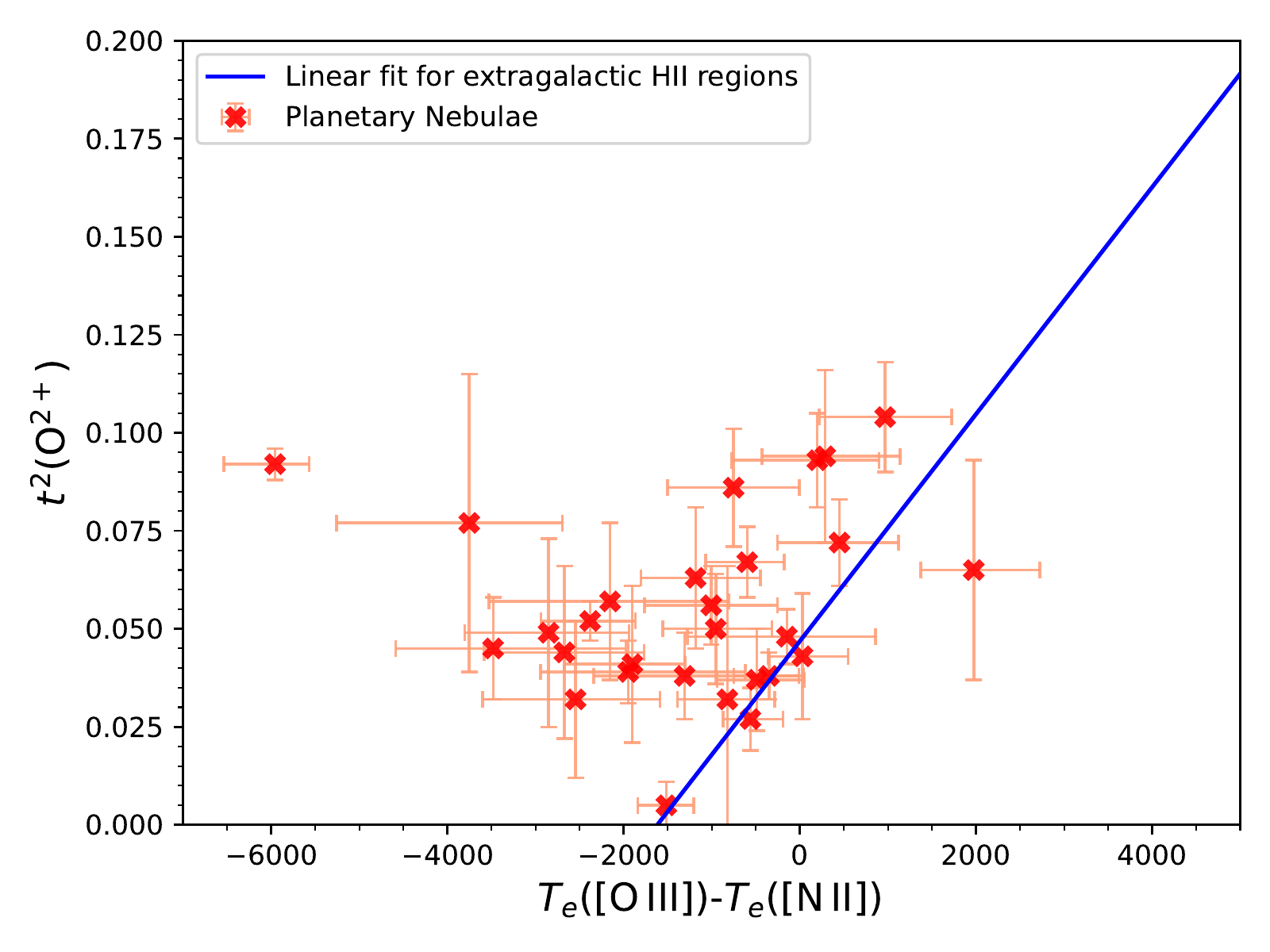}
\caption{$t^2(\text{O}^{2+})$ versus $\Delta T_{\rm e}$ relation for Galactic planetary nebulae.}\label{Fig:Delta_t2-3}
\end{figure}

Finally, we explore the effect of temperature inhomogeneities on the PNe shown in Table~\ref{tab:densities_samplePNe} following the same methodology described for H\,II regions. Fig.~\ref{Fig:Delta_t2-3} shows that, in general, PNe do not follow the same trend between $t^2(\text{O}^{2+})$ and $\Delta T_{\rm e}$ as H\,II regions. It should be noted that the ionizing sources of these objects are hotter, having a considerable fraction of $\text{O}^{3+}/\text{O}$. Therefore the heating may be affecting mainly the $\text{O}^{3+}$ volume, which is not accessible in the optical range. Furthermore, there are strong indications that some PNe with binary central stars can have inclusions of (at least) two distinct gas components with different metallicities \cite{Liu:2001,Wesson2003,Liu2006,Storey2014,garciarojas2022}. These chemical inhomogeneities also produce temperature variations, but they imply the invalidity of $t^2(\text{O}^{2+})$ derived from $T_{\rm e}$(O\,II V1/[O\,III]$\lambda 5007$) as the emission of RLs and CELs would come from non related volumes \cite{Zhang:2007}. This is particularly important in the extreme ADF PNe Ou5 and Abell\,46 \cite{corradi15}. The abundance discrepancy problem in PNe remains open and requires further investigations, although \cite{Richer2022} found that the effect of temperature inhomogeneities on these objects is not negligible .

\begin{table}[h]
\begin{center}
\begin{minipage}{\textwidth}
\caption{Adopted atomic data set.}\label{tab:atomic_data}
\begin{tabular*}{\textwidth}{@{\extracolsep{\fill}}lccccccccccccc@{\extracolsep{\fill}}}
\toprule%
\multicolumn{1}{l}{Ion} & \multicolumn{1}{c}{Transition Probabilities} &
\multicolumn{1}{c}{Collision Strengths}&
\multicolumn{1}{c}{Effective recombination coefficients} \\
\midrule
H$^{+}$   & - & -& \cite{Storey95}\\
O$^{+}$   &  \cite{FroeseFischer:2004} & \cite{Kisielius:2009}&-\\
O$^{2+}$  &  \cite{Wiese:1996}, \cite{Storey:2000} & \cite{Storey:2014}& \cite{Storey2017}\\
N$^{+}$   &  \cite{FroeseFischer:2004} & \cite{Tayal:2011}&-\\
S$^{+}$   &  \cite{IFF05} & \cite{Tayal:2010}&-\\
Cl$^{2+}$ &  \cite{Fritzsche:1999} & \cite{Butler:1989}&-\\
Ar$^{2+}$ &   \cite{Mendoza:1983b}, \cite{Kaufman:1986}  & \cite{Galavis:1995}&-\\
Ar$^{3+}$ &   \cite{Mendoza:1982b}  & \cite{Ramsbottom:1997}&-\\
Fe$^{2+}$ & \cite{Quinet:1996} & \cite{Zhang:1996}&-\\
\botrule
\end{tabular*}
\end{minipage}
\end{center}
\end{table}

\backmatter

\bmhead{Acknowledgments}

JEM-D appreciates the fruitful conversations with Antonio Peimbert and Silvia Torres-Peimbert on the formalism of temperature variations and chemical inhomogeneities in the ionized gas. JEM-D acknowledges interesting discussions with William J. Henney.

\bmhead{Authors' contributions}

JEM-D conducted the study with the original idea, compiled the appropriate data, re-calculated the physical conditions and chemical abundances, wrote the text, created the figures and interpreted the results. CE, JG-R, KK and MP reviewed the consistency of the analysis, the formalism of temperature fluctuations, contributed to data interpretation and discussion, and modified the text. The participation of all the authors was essential for the realization of this work.

\bmhead{Conflict of interest/Competing interests}

The authors declare that they have no competing financial interests.

\bmhead{Data availability}

All the data is public and available in cited the references. Our calculations are entirely present in Tables~\ref{tab:physical_cond_sampleextra}-\ref{tab:ionicabundances}. 

\bmhead{Funding}

JEM-D and KK gratefully acknowledge funding from the Deutsche Forschungsgemeinschaft (DFG, German Research Foundation) in the form of an Emmy Noether Research Group (grant number KR4598/2-1, PI Kreckel).  CE and JG-R acknowledge support from the Agencia Estatal de Investigaci\'on del Ministerio de Ciencia e Innovaci\'on (AEI-MCINN) under grant {\it Espectroscop\'ia de campo integral de regiones H\,II locales. Modelos para el estudio de regiones H\,II extragal\'acticas} with reference 10.13039/501100011033 and support under grant P/308614 financed by funds transferred from the Spanish Ministry of Science, Innovation and Universities, charged to the General State Budgets and with funds transferred from the General Budgets of the Autonomous Community of the Canary Islands by the MCIU. JG-R acknowledges support from an Advanced Fellowship under the Severo Ochoa excellence program CEX2019-000920-S and financial support from the Canarian Agency for Research, Innovation and Information Society (ACIISI), of the Canary Islands Government, and the European Regional Development Fund (ERDF), under grant with reference ProID2021010074.

\bmhead{Additional Information}

Correspondence should be addressed to JEM-D: jemd@uni-heidelberg.de

\begin{appendices}

\section{Physical conditions, ionic abundances and references}\label{secA1}

\begin{sidewaystable}
\sidewaystablefn%
\begin{center}
\begin{minipage}{\textheight}
\caption{Physical conditions for our sample of extragalactic H\,II regions.}\label{tab:physical_cond_sampleextra}
\begin{tabular*}{\textheight}{@{\extracolsep{\fill}}lccccccccccccc@{\extracolsep{\fill}}}
\toprule%
Region & $n_{\rm e}$([O\,II])  & $n_{\rm e}$([S\,II]) & Adop. $n_{\rm e}$ & $T_{\rm e}$([O\,III])& $T_{\rm e}$([Ar\,III])& $T_{\rm e}$([N\,II]) &Ref.\\
 & cm$^{-3}$ & cm$^{-3}$ & cm$^{-3}$ & K & K & K & \\

\midrule
NGC\,5461 & $310^{+90} _{-80}$ & $220^{+110} _{-120}$ & $270 \pm 100$ & $8460^{+200} _{-220}$ & - & $9100^{+300} _{-350}$ & \cite{esteban09}\\
NGC\,588 & $90^{+110} _{-70}$ & $150^{+120} _{-110}$ & $120^{+110} _{-90}$ & $10810^{+500} _{-380}$ & - & $10900^{+750} _{-1200}$ & \cite{Toribio16}\\
NGC\,595 & $60 \pm 20$ & $40 \pm 30$ & $<100$ & $7390^{+390} _{-400}$ & - & $8150^{+210} _{-140}$ & \cite{esteban09}\\
K\,932 & $250 \pm 30$ & $110 \pm 50$ & $180 \pm 40$ & $8300^{+140} _{-130}$ & - & $9260^{+170} _{-190}$ & \cite{esteban09}\\
IC\,132 & $160^{+200} _{-110}$ & $200^{+170} _{-120}$ & $180^{+180} _{-120}$ & $10280^{+450} _{-540}$ & $12020^{+1000} _{-1150}$ & $11300^{+1050} _{-980}$ & \cite{Toribio16}\\
N\,11B & $290 \pm 100$ & $260^{+60} _{-50}$ & $270 \pm 80$ & $9140^{+130} _{-90}$ & $9680^{+260} _{-190}$ & $9780^{+160} _{-150}$ & \cite{dominguezguzman22}\\
NGC\,5471 & $210^{+60} _{-50}$ & $170^{+70} _{-60}$ & $190^{+60} _{-50}$ & $13820^{+190} _{-150}$ & $14520^{+1220} _{-1130}$ & $12180^{+540} _{-370}$ & \cite{esteban20}\\
N\,66A & $180^{+100} _{-90}$ & $180 \pm 70$ & $180 \pm 80$ & $12550 \pm 180$ & $12220^{+690} _{-640}$ & $12010^{+330} _{-290}$ & \cite{dominguezguzman22}\\
NGC\,604 & $60 \pm 20$ & $50^{+40} _{-30}$ & $<100$ & $8050^{+140} _{-200}$ & - & $8640^{+260} _{-240}$ & \cite{esteban09}\\
NGC\,6822 & $150^{+100} _{-90}$ & $110^{+90} _{-70}$ & $130^{+100} _{-80}$ & $11510^{+190} _{-260}$ & $11800^{+1330} _{-1420}$ & $11340^{+1290} _{-1720}$ & \cite{esteban14}\\
UV-1 & $190 \pm 70$ & $280 \pm 80$ & $240 \pm 80$ & $10840^{+190} _{-130}$ & $10610^{+1000} _{-1310}$ & $10130^{+520} _{-710}$ & \cite{lopezsanchez07}\\
NGC\,5408 & $220 \pm 90$ & $200^{+110} _{-90}$ & $210^{+100} _{-90}$ & $15910^{+330} _{-460}$ & - & $11490^{+790} _{-800}$ & \cite{esteban14}\\
NGC\,1714 & $450^{+140} _{-130}$ & $340^{+140} _{-130}$ & $390^{+140} _{-130}$ & $9520^{+170} _{-220}$ & $9430^{+700} _{-770}$ & $10190^{+560} _{-420}$ & \cite{dominguezguzman22}\\
N\,81 & $440 \pm 130$ & $340^{+80} _{-90}$ & $390^{+100} _{-110}$ & $12830^{+220} _{-250}$ & $11870^{+190} _{-270}$ & $11850^{+220} _{-280}$ & \cite{dominguezguzman22}\\
IC\,2111 & $290 \pm 120$ & $240^{+120} _{-100}$ & $270^{+120} _{-110}$ & $9130^{+180} _{-200}$ & $9120^{+480} _{-740}$ & $9760^{+360} _{-310}$ & \cite{dominguezguzman22}\\
NGC\,2363 & $270^{+90} _{-100}$ & $170^{+120} _{-90}$ & $220^{+110} _{-100}$ & $16480^{+450} _{-390}$ & $15600^{+1200} _{-1100}$ & $12970^{+1220} _{-1080}$ & \cite{esteban09}\\
30Dor & $440 \pm 20$ & $350 \pm 20$ & $390 \pm 20$ & $9870^{+90} _{-40}$ & $9120^{+400} _{-300}$ & $10360^{+280} _{-240}$ & \cite{peimbert03}\\
N\,44C & $190^{+110} _{-80}$ & $100 \pm 50$ & $140^{+80} _{-70}$ & $11280^{+150} _{-140}$ & $10630 \pm 530$ & $10560^{+320} _{-330}$ & \cite{dominguezguzman22}\\
HII-2 & $420^{+110} _{-90}$ & $380^{+100} _{-90}$ & $400^{+110} _{-90}$ & $11900^{+240} _{-190}$ & $12230^{+1210} _{-1300}$ & $10520^{+500} _{-530}$ & \cite{lopezsanchez07}\\
NGC\,5455 & $240^{+60} _{-50}$ & $160 \pm 50$ & $200 \pm 50$ & $9610^{+120} _{-150}$ & $10350^{+1120} _{-900}$ & $10120^{+200} _{-170}$ & \cite{esteban20}\\
VS44 & $150 \pm 20$ & $100^{+30} _{-40}$ & $<100$ & $8210^{+240} _{-170}$ & - & $8870^{+330} _{-250}$ & \cite{esteban09}\\
N\,88A & $2700^{+570} _{-400}$ & $1840^{+400} _{-310}$ & $3130 \pm 1120$ & $13820^{+250} _{-190}$ & $20190^{+680} _{-670}$ & $13020^{+280} _{-310}$ & \cite{dominguezguzman22}\\
HII-1 & $360 \pm 90$ & $420^{+110} _{-90}$ & $390^{+100} _{-90}$ & $11950^{+240} _{-160}$ & $12210^{+1010} _{-1220}$ & $10810^{+460} _{-520}$ & \cite{lopezsanchez07}\\

\botrule
\end{tabular*}
\end{minipage}
\end{center}
\end{sidewaystable}

\begin{table}[h]
\begin{center}
\begin{minipage}{\textwidth}
\caption{Physical conditions of the adopted sample of Galactic ring nebulae.}\label{tab:physical_cond_sampleRNe}
\begin{tabular*}{\textwidth}{@{\extracolsep{\fill}}lcccccc@{\extracolsep{\fill}}}
\toprule%

Region &  $n_{\rm e}$([S\,II]) & $T_{\rm e}$([O\,III])& $T_{\rm e}$([Ar\,III])& $T_{\rm e}$([N\,II]) &Ref.\\
 & cm$^{-3}$ & K & K & K & \\

\midrule
NGC\,7635A2 & $120 \pm 40$ & $8130^{+400} _{-550}$ & - & $8000^{+560} _{-600}$ & \cite{esteban16}\\
NGC\,7635A3 & $90^{+50} _{-40}$ & $7530^{+610} _{-520}$ & - & $8190^{+420} _{-300}$ & \cite{esteban16}\\
NGC\,7635A4 & $1450^{+130} _{-140}$ & $8650^{+350} _{-480}$ & $8410^{+720} _{-810}$ & $9010^{+110} _{-90}$ & \cite{esteban16}\\
NGC\,6888A2 & $290 \pm 80$ & $12710^{+400} _{-410}$ & - & $7380 \pm 110$ & \cite{esteban16}\\
NGC\,6888A3 & $210^{+100} _{-80}$ & $10040^{+500} _{-870}$ & - & $7570^{+160} _{-180}$ & \cite{esteban16}\\
NGC\,6888A4 & $120^{+80} _{-60}$ & $10050^{+820} _{-930}$ & - & $8800^{+200} _{-110}$ & \cite{esteban16}\\
NGC\,6888A5 & $170 \pm 70$ & $9850^{+420} _{-580}$ & $7310^{+740} _{-610}$ & $8660^{+220} _{-240}$ & \cite{esteban16}\\
NGC\,6888A6 & $180^{+100} _{-90}$ & $9620^{+620} _{-870}$ & - & $8660^{+210} _{-160}$ & \cite{esteban16}\\

\botrule
\end{tabular*}
\end{minipage}
\end{center}
\end{table}

\begin{sidewaystable}
\sidewaystablefn%
\begin{center}
\begin{minipage}{\textheight}
\caption{Physical conditions of the extragalactic H\,II regions from other research groups.}\label{tab:physical_cond_sampleextra_others}
\begin{tabular*}{\textheight}{@{\extracolsep{\fill}}lccccccccccccc@{\extracolsep{\fill}}}
\toprule%
Region & $n_{\rm e}$([O\,II])  & $n_{\rm e}$([S\,II]) & Adop. $n_{\rm e}$ & $T_{\rm e}$([O\,III])& $T_{\rm e}$([Ar\,III])& $T_{\rm e}$([N\,II]) &Ref.\\
 & cm$^{-3}$ & cm$^{-3}$ & cm$^{-3}$ & K & K & K & \\

\midrule
NGC\,456a-1 & - & $290^{+40} _{-50}$ & $290^{+40} _{-50}$ & $12210^{+70} _{-130}$ & - & $11890^{+470} _{-380}$ & \cite{guseva11}\\
NGC\,456a-2 & - & $80^{+40} _{-30}$ & $<100$ & $11940^{+70} _{-100}$ & - & $10700^{+980} _{-760}$ & \cite{guseva11}\\
NGC\,6822V & $70^{+50} _{-30}$ & $90^{+70} _{-50}$ & $<100$ & $11700^{+280} _{-210}$ & - & $14520^{+1970} _{-2280}$ & \cite{peimbert05}\\
NGC\,5253C1 & - & $530^{+70} _{-50}$ & $530^{+70} _{-50}$ & $12140^{+80} _{-70}$ & - & $11420^{+100} _{-150}$ & \cite{guseva11}\\
Mrk1259 & - & $780^{+80} _{-70}$ & $780^{+80} _{-70}$ & $9970^{+460} _{-260}$ & - & $7860^{+250} _{-280}$ & \cite{guseva11}\\
NGC\,5253C2 & - & $200 \pm 40$ & $200 \pm 40$ & $10150^{+80} _{-50}$ & - & $11010^{+440} _{-350}$ & \cite{guseva11}\\
NGC\,5253P2 & - & $810 \pm 80$ & $810 \pm 80$ & $12280^{+120} _{-110}$ & - & $11320^{+220} _{-200}$ & \cite{guseva11}\\
NGC\,346 & $40^{+30} _{-20}$ & $40^{+30} _{-20}$ & $<100$ & $12820^{+120} _{-110}$ & - & $10830^{+860} _{-880}$ & \cite{valerdi19}\\
NGC\,456-2 & $210 \pm 20$ & $210 \pm 50$ & $210 \pm 40$ & $12070^{+170} _{-220}$ & $11320^{+1250} _{-1150}$ & $11110^{+1070} _{-760}$ & \cite{penaguerrero12}\\

\botrule
\end{tabular*}
\end{minipage}
\end{center}
\end{sidewaystable}

\begin{sidewaystable}
\sidewaystablefn%
\begin{center}
\begin{minipage}{\textheight}
\caption{Electron density values of the adopted sample of Galactic H\,II regions.}\label{tab:densities_sampleGal}
\begin{tabular*}{\textheight}{@{\extracolsep{\fill}}lccccccccccccc@{\extracolsep{\fill}}}
\toprule%
Region & $n_{\rm e}$([O\,II])  & $n_{\rm e}$([S\,II]) & $n_{\rm e}$([Fe\,III])  & $n_{\rm e}$([Cl\,III]) &  $n_{\rm e}$([Ar\,IV]) & Adop. $n_{\rm e}$ &Ref.\\
 & cm$^{-3}$ & cm$^{-3}$ & cm$^{-3}$ & cm$^{-3}$ & cm$^{-3}$ & cm$^{-3}$ & \\
 
\midrule

NGC\,3603 & $2600^{+800} _{-510}$ & $2890^{+1140} _{-650}$ & $8400^{+28360} _{-5500}$ & $4850^{+1350} _{-1570}$ & $1950^{+1620} _{-1210}$ & $2850 \pm 750$ & \cite{garciarojas06}\\
M\,42-P1 & $4700^{+950} _{-870}$ & $2870^{+670} _{-600}$ & $6590^{+3200} _{-2010}$ & $5930^{+1380} _{-1020}$ & $5920^{+3800} _{-3310}$ & $3990 \pm 1280$ & \cite{delgadoinglada16}\\
M\,42-2 & $5850^{+820} _{-770}$ & $3760^{+1390} _{-750}$ & $10340^{+2180} _{-2130}$ & $7510^{+820} _{-640}$ & $6690^{+580} _{-480}$ & $6500 \pm 1240$ & \cite{mendezdelgado22b}\\
Sh\,2-100 & - & $390^{+230} _{-170}$ & $13340^{+21970} _{-9720}$ & $750^{+410} _{-330}$ & - & $390^{+230} _{-170}$ & \cite{esteban17}\\
M\,42-3 & $5320^{+710} _{-550}$ & $4100^{+1160} _{-950}$ & $11300^{+3410} _{-2330}$ & $7310^{+900} _{-830}$ & $5950^{+1410} _{-1170}$ & $5810 \pm 1350$ & \cite{mendezdelgado21a}\\
M\,42-1 & $5340^{+720} _{-560}$ & $4180^{+1020} _{-800}$ & $10310^{+2850} _{-2670}$ & $6900^{+720} _{-610}$ & $4190^{+1380} _{-1270}$ & $5670 \pm 1260$ & \cite{mendezdelgado21a}\\
Sh\,2-311 & $270^{+80} _{-60}$ & $300 \pm 100$ & $6420^{+17160} _{-4890}$ & - & - & $290^{+90} _{-80}$ & \cite{garciarojas05}\\
M\,16 & $1160^{+210} _{-190}$ & $1080^{+220} _{-190}$ & $10160^{+20300} _{-8180}$ & $1190^{+530} _{-510}$ & - & $1130 \pm 100$ & \cite{garciarojas06}\\
M\,42-bar & $4200^{+490} _{-410}$ & $3240^{+750} _{-570}$ & $4240^{+2010} _{-1790}$ & $4330^{+550} _{-640}$ & $30870^{+48400} _{-23610}$ & $4030 \pm 480$ & \cite{delgadoinglada16}\\
M\,42-1 & $1110^{+80} _{-90}$ & $1320^{+190} _{-170}$ & $4120^{+1790} _{-1260}$ & $1540^{+230} _{-240}$ & - & $1190 \pm 190$ & \cite{mendezdelgado21b}\\
NGC\,3576 & $1710^{+260} _{-280}$ & $1050^{+300} _{-270}$ & $1560^{+1350} _{-1380}$ & $2890^{+630} _{-620}$ & $3050^{+1620} _{-1380}$ & $1560 \pm 550$ & \cite{garciarojas04}\\
M\,42 & $6790^{+2510} _{-1620}$ & $4550^{+3150} _{-1500}$ & $9550^{+3570} _{-2820}$ & $7020^{+610} _{-540}$ & $4810^{+1190} _{-900}$ & $6510 \pm 1060$ & \cite{esteban04}\\
M\,17 & $500^{+110} _{-100}$ & $380 \pm 110$ & $9560^{+32890} _{-7330}$ & $370^{+270} _{-200}$ & - & $440^{+110} _{-100}$ & \cite{garciarojas07}\\
M\,8 & $1630^{+670} _{-500}$ & $1230^{+210} _{-170}$ & $2440^{+1810} _{-1130}$ & $1870^{+250} _{-270}$ & $4140^{+5230} _{-2800}$ & $1480 \pm 330$ & \cite{garciarojas07}\\
M\,42-4 & $4860^{+620} _{-510}$ & $3980^{+790} _{-660}$ & $9660^{+3090} _{-2230}$ & $6970^{+670} _{-630}$ & $5220^{+820} _{-710}$ & $5360 \pm 1200$ & \cite{mendezdelgado21a}\\
M\,42-3 & $5350^{+720} _{-630}$ & $3950^{+860} _{-570}$ & $7580^{+2590} _{-1950}$ & $7650^{+810} _{-790}$ & $4670^{+500} _{-410}$ & $5170 \pm 1180$ & \cite{mendezdelgado22b}\\
Sh\,2-288 & - & $430^{+250} _{-230}$ & $2960^{+6080} _{-2240}$ & $440^{+360} _{-270}$ & - & $430^{+250} _{-230}$ & \cite{esteban17}\\
M\,42 & $2710^{+690} _{-540}$ & $1800^{+570} _{-520}$ & $4370^{+6120} _{-3020}$ & $2120^{+970} _{-800}$ & $89060^{+317740} _{-63110}$ & $2200 \pm 480$ & \cite{mesadelgado09}\\
M\,42-2 & $1460^{+140} _{-110}$ & $1160^{+190} _{-140}$ & $3530^{+2340} _{-1490}$ & $2000^{+500} _{-470}$ & - & $1380 \pm 220$ & \cite{mendezdelgado21b}\\
M\,42-2 & $5110^{+660} _{-560}$ & $4090^{+890} _{-770}$ & $10160^{+2490} _{-1990}$ & $6510^{+650} _{-610}$ & $5870 \pm 720$ & $5620 \pm 1080$ & \cite{mendezdelgado21a}\\

\botrule
\end{tabular*}
\end{minipage}
\end{center}
\end{sidewaystable}

\begin{sidewaystable}
\sidewaystablefn%
\begin{center}
\begin{minipage}{\textheight}
\caption{Electron density values of the adopted sample of Galactic planetary nebulae.}\label{tab:densities_samplePNe}
\begin{tabular*}{\textheight}{@{\extracolsep{\fill}}lccccccccccccc@{\extracolsep{\fill}}}
\toprule%
Region & $n_{\rm e}$([O\,II])  & $n_{\rm e}$([S\,II]) & $n_{\rm e}$([Fe\,III])  & $n_{\rm e}$([Cl\,III]) &  $n_{\rm e}$([Ar\,IV]) & Adop. $n_{\rm e}$ &Ref.\\
 & cm$^{-3}$ & cm$^{-3}$ & cm$^{-3}$ & cm$^{-3}$ & cm$^{-3}$ & cm$^{-3}$ & \\
 
\midrule
NGC\,2440 & $4080^{+620} _{-590}$ & $2620^{+420} _{-380}$ & - & $5520^{+550} _{-510}$ & $5000^{+610} _{-600}$ & $3990 \pm 1220$ & \cite{sharpee07}\\
Hb\,4 & $7200^{+13470} _{-3200}$ & $5200^{+5010} _{-2780}$ & - & $6930^{+1680} _{-1750}$ & $7320^{+1190} _{-1380}$ & $7060 \pm 520$ & \cite{garciarojas12}\\
NGC\,5315 & $12510^{+22180} _{-6340}$ & $7920^{+11590} _{-4380}$ & $10730^{+19690} _{-7230}$ & $34670^{+11420} _{-9530}$ & $37220^{+10270} _{-8950}$ & $20930 \pm 13320$ & \cite{madonna17}\\
M\,1-31 & $15130^{+29210} _{-7920}$ & $9990^{+7630} _{-4110}$ & - & $19820^{+14290} _{-9620}$ & - & $12140 \pm 3840$ & \cite{garciarojas18}\\
Cn\,1-5 & $5070^{+2560} _{-1490}$ & $3930^{+1980} _{-1290}$ & $14260^{+21830} _{-9390}$ & $3910^{+860} _{-1010}$ & $10950^{+6910} _{-6020}$ & $4190 \pm 1010$ & \cite{garciarojas12}\\
H\,1-40 & $7790^{+18670} _{-4100}$ & $8300^{+8690} _{-3460}$ & $16750^{+36490} _{-12980}$ & $7630^{+7420} _{-5100}$ & - & $8180 \pm 1450$ & \cite{garciarojas18}\\
H\,1-50 & $10670^{+8880} _{-3730}$ & $7160^{+4300} _{-2650}$ & - & $12850^{+4170} _{-3710}$ & - & $9790 \pm 2630$ & \cite{garciarojas18}\\
Hen\,2-158 & $3780^{+1180} _{-760}$ & $2750^{+770} _{-670}$ & $12130^{+30230} _{-9990}$ & $6470^{+4100} _{-3620}$ & $18870^{+22320} _{-12970}$ & $3220 \pm 890$ & \cite{garciarojas18}\\
IC\,418 & $15380^{+8370} _{-4520}$ & $14210^{+9850} _{-5610}$ & - & $12200^{+1150} _{-1210}$ & $5130^{+4020} _{-2990}$ & $11640 \pm 2220$ & \cite{sharpee03}\\
PC\,14 & $4520^{+1630} _{-1300}$ & $3070^{+1640} _{-960}$ & - & $3410^{+860} _{-830}$ & $4710^{+1120} _{-1140}$ & $3820 \pm 660$ & \cite{garciarojas12}\\
M\,1-32 & $9570^{+14360} _{-5220}$ & $5940^{+12420} _{-3190}$ & $20710^{+20290} _{-10300}$ & $13460^{+3620} _{-3110}$ & - & $12350 \pm 3120$ & \cite{garciarojas12}\\
M\,1-61 & $19180^{+25650} _{-9700}$ & $9820^{+14210} _{-5100}$ & $20290^{+34130} _{-16340}$ & $19550^{+5420} _{-3510}$ & $33540^{+9600} _{-6570}$ & $20900 \pm 6910$ & \cite{garciarojas12}\\
Ou5 & $1380^{+510} _{-390}$ & $220^{+220} _{-150}$ & - & - & $1200^{+1500} _{-820}$ & $800^{+360} _{-270}$ & \cite{corradi15}\\
NGC\,6369 & $3940^{+1540} _{-1060}$ & $3170^{+1530} _{-1000}$ & - & $4160^{+970} _{-1010}$ & $4960^{+1410} _{-1160}$ & $4070 \pm 600$ & \cite{garciarojas12}\\
Pe\,1-1 & $21170^{+33240} _{-11750}$ & $9850^{+9830} _{-4670}$ & - & $31030^{+12290} _{-8480}$ & - & $17080 \pm 9670$ & \cite{garciarojas12}\\
IC\,2501 & $16950^{+8770} _{-5170}$ & $8340^{+3510} _{-1910}$ & - & $9970^{+990} _{-960}$ & $8490^{+770} _{-660}$ & $9030 \pm 970$ & \cite{sharpee07}\\
M\,1-30 & $4480^{+3020} _{-1550}$ & $5160^{+6130} _{-2420}$ & $9530^{+17230} _{-7780}$ & $7380^{+1240} _{-1450}$ & - & $6560 \pm 1300$ & \cite{garciarojas12}\\
IC\,4191 & $14150^{+5370} _{-3740}$ & $7620^{+2830} _{-1930}$ & - & $14290^{+1550} _{-1090}$ & $11310^{+1090} _{-980}$ & $11990 \pm 2020$ & \cite{sharpee07}\\
Hen\,2-96 & $13340^{+23340} _{-6310}$ & $8980^{+6130} _{-3290}$ & - & $23960^{+27980} _{-13770}$ & $38610^{+28930} _{-18170}$ & $10980 \pm 6050$ & \cite{garciarojas18}\\
M\,3-15 & $9780^{+27490} _{-4980}$ & $5260^{+6820} _{-2970}$ & - & $9260^{+2760} _{-2190}$ & $8270^{+4480} _{-4220}$ & $8430 \pm 1440$ & \cite{garciarojas12}\\
Hen\,2-86 & $11550^{+20490} _{-5240}$ & $9010^{+12200} _{-4470}$ & $41590^{+33180} _{-19220}$ & $21620^{+6190} _{-3980}$ & $36560^{+5390} _{-5470}$ & $24730 \pm 10360$ & \cite{garciarojas12}\\
M\,2-31 & $7560^{+6610} _{-2410}$ & $4740^{+2150} _{-1330}$ & - & $9990^{+3790} _{-3290}$ & $8080^{+1670} _{-1510}$ & $6940 \pm 1830$ & \cite{garciarojas18}\\
NGC\,5189 & $1270^{+380} _{-350}$ & $1000^{+450} _{-380}$ & - & $1370^{+550} _{-560}$ & $1240^{+650} _{-610}$ & $1130^{+410} _{-370}$ & \cite{garciarojas12}\\
M\,1-60 & $10590^{+15100} _{-4300}$ & $6170^{+3220} _{-1900}$ & - & $14110^{+8650} _{-5700}$ & $18230^{+2660} _{-3580}$ & $11210 \pm 5650$ & \cite{garciarojas18}\\
Hen\,2-73 & $9100^{+5880} _{-3010}$ & $6510^{+4510} _{-2260}$ & - & $12660^{+4620} _{-3540}$ & - & $9040 \pm 2610$ & \cite{garciarojas18}\\
IC\,4776 & $14320^{+23970} _{-7380}$ & $11380^{+9520} _{-4400}$ & $19480^{+27120} _{-10010}$ & $23470^{+6770} _{-6590}$ & $36820^{+3160} _{-3480}$ & $29950 \pm 9850$ & \cite{sowicka17}\\
M\,1-25 & $12300^{+17390} _{-5210}$ & $6970^{+11720} _{-3690}$ & $11010^{+20430} _{-8120}$ & $13690^{+2900} _{-2250}$ & - & $12930 \pm 1980$ & \cite{garciarojas12}\\
M\,1-33 & $5550^{+3300} _{-1390}$ & $4370^{+1750} _{-1250}$ & $14960^{+55300} _{-13430}$ & $7230^{+2650} _{-2370}$ & - & $5230 \pm 1160$ & \cite{garciarojas18}\\
M\,2-36 & $3900^{+490} _{-450}$ & $2660^{+300} _{-270}$ & - & $5190^{+1130} _{-1200}$ & $3960^{+930} _{-870}$ & $3140 \pm 710$ & \cite{espiritu21}\\
Abell\,46 & $3210^{+1060} _{-710}$ & $440^{+400} _{-310}$ & - & - & $1370^{+1630} _{-890}$ & $1820^{+730} _{-510}$ & \cite{corradi15}\\

\botrule
\end{tabular*}
\end{minipage}
\end{center}
\end{sidewaystable}

\begin{table}[h]
\begin{center}
\begin{minipage}{\textwidth}
\caption{Electron temperature values of the adopted sample of Galactic H\,II regions and planetary nebulae.}\label{tab:temperatures_sampleGalPNe}
\begin{tabular*}{\textwidth}{@{\extracolsep{\fill}}lcccccc@{\extracolsep{\fill}}}
\toprule%

Region & $T_{\rm e}$([O\,III])& $T_{\rm e}$([Ar\,III])& $T_{\rm e}$([N\,II]) &Ref.\\
 & K & K & K &    \\
\midrule
\multicolumn{5}{c}{{\bf Galactic H\,II regions}}\\
\\
NGC\,3603 & $9020^{+130} _{-120}$ & - & $11040^{+650} _{-530}$ & \cite{garciarojas06}\\
M\,42-P1 & $8320^{+80} _{-60}$ & $7970 \pm 440$ & $9500^{+250} _{-260}$ & \cite{delgadoinglada16}\\
M\,42-2 & $8500^{+60} _{-70}$ & $8100^{+130} _{-150}$ & $9810^{+230} _{-200}$ & \cite{mendezdelgado22b}\\
Sh\,2-100 & $8280^{+100} _{-120}$ & $7650^{+380} _{-520}$ & $8430^{+270} _{-300}$ & \cite{esteban17}\\
M\,42-3 & $8510 \pm 60$ & $8260^{+160} _{-200}$ & $10090^{+230} _{-300}$ & \cite{mendezdelgado21a}\\
M\,42-1 & $8420^{+80} _{-60}$ & $8270^{+210} _{-220}$ & $9920^{+210} _{-200}$ & \cite{mendezdelgado21a}\\
Sh\,2-311 & $8960^{+90} _{-100}$ & $8830^{+890} _{-950}$ & $9230^{+170} _{-240}$ & \cite{garciarojas05}\\
M\,16 & $7600^{+140} _{-130}$ & - & $8280^{+230} _{-130}$ & \cite{garciarojas06}\\
M\,42-bar & $8470^{+90} _{-70}$ & $8330^{+320} _{-290}$ & $9000^{+130} _{-150}$ & \cite{delgadoinglada16}\\
M\,42-1 & $8020^{+60} _{-50}$ & - & $8510^{+140} _{-110}$ & \cite{mendezdelgado21b}\\
NGC\,3576 & $8450^{+40} _{-50}$ & $8590^{+340} _{-270}$ & $8710^{+160} _{-270}$ & \cite{garciarojas04}\\
M\,42 & $8380 \pm 60$ & $8190^{+250} _{-360}$ & $9920^{+290} _{-220}$ & \cite{esteban04}\\
M\,17 & $7950^{+100} _{-90}$ & $8350^{+480} _{-510}$ & $8740^{+220} _{-180}$ & \cite{garciarojas07}\\
M\,8 & $8050^{+110} _{-100}$ & $7570 \pm 310$ & $8330 \pm 100$ & \cite{garciarojas07}\\
M\,42-4 & $8320^{+90} _{-60}$ & $8250^{+320} _{-360}$ & $9890^{+220} _{-210}$ & \cite{mendezdelgado21a}\\
M\,42-3 & $8410^{+50} _{-60}$ & $8310 \pm 200$ & $9890^{+190} _{-260}$ & \cite{mendezdelgado22b}\\
Sh\,2-288 & $9260^{+520} _{-620}$ & - & $9390^{+340} _{-350}$ & \cite{esteban17}\\
M\,42 & $8180^{+140} _{-260}$ & $7880^{+460} _{-520}$ & $9430^{+380} _{-320}$ & \cite{mesadelgado09}\\
M\,42-2 & $8120^{+50} _{-70}$ & - & $8450^{+140} _{-150}$ & \cite{mendezdelgado21b}\\
M\,42-2 & $8390^{+70} _{-50}$ & $8330^{+180} _{-130}$ & $9860^{+230} _{-170}$ & \cite{mendezdelgado21a}\\
\\
\multicolumn{5}{c}{{\bf Galactic planetary nebulae}}\\
\\
NGC\,2440 & $14850^{+290} _{-300}$ & $12120 \pm 260$ & $13880^{+470} _{-450}$ & \cite{sharpee07}\\
Hb\,4 & $10010^{+260} _{-290}$ & $8290^{+550} _{-400}$ & $9810^{+440} _{-690}$ & \cite{garciarojas12}\\
NGC\,5315 & $8760^{+220} _{-250}$ & $8330^{+280} _{-320}$ & $10060^{+1130} _{-780}$ & \cite{madonna17}\\
M\,1-31 & $8310^{+180} _{-240}$ & - & $10210^{+410} _{-530}$ & \cite{garciarojas18}\\
Cn\,1-5 & $8740 \pm 180$ & $7840^{+320} _{-330}$ & $8710^{+340} _{-210}$ & \cite{garciarojas12}\\
H\,1-40 & $10100^{+390} _{-370}$ & - & $13850^{+670} _{-1140}$ & \cite{garciarojas18}\\
H\,1-50 & $11090 \pm 230$ & $10480 \pm 820$ & $11840^{+510} _{-520}$ & \cite{garciarojas18}\\
Hen\,2-158 & $9280^{+180} _{-220}$ & - & $10100 \pm 350$ & \cite{garciarojas18}\\
IC\,418 & $8770^{+90} _{-100}$ & $8600^{+380} _{-300}$ & $9340^{+280} _{-210}$ & \cite{sharpee03}\\
PC\,14 & $9270^{+210} _{-170}$ & $8830^{+440} _{-550}$ & $10220 \pm 430$ & \cite{garciarojas12}\\
M\,1-32 & $9370^{+240} _{-210}$ & $8040^{+620} _{-800}$ & $8920^{+430} _{-490}$ & \cite{garciarojas12}\\
M\,1-61 & $9190^{+200} _{-240}$ & $8730^{+380} _{-370}$ & $12680^{+1310} _{-870}$ & \cite{garciarojas12}\\
Ou5 & $12470^{+440} _{-380}$ & - & $25870^{+9940} _{-6420}$ & \cite{corradi15}\\
NGC\,6369 & $10650^{+330} _{-240}$ & $9280^{+800} _{-920}$ & $13320^{+570} _{-670}$ & \cite{garciarojas12}\\
Pe\,1-1 & $10100^{+340} _{-290}$ & $9220^{+550} _{-490}$ & $12250^{+1010} _{-1080}$ & \cite{garciarojas12}\\
IC\,2501 & $9410^{+80} _{-110}$ & $8840^{+320} _{-250}$ & $10930^{+230} _{-210}$ & \cite{sharpee07}\\
M\,1-30 & $6640^{+80} _{-170}$ & $6430^{+180} _{-250}$ & $6990^{+260} _{-230}$ & \cite{garciarojas12}\\
IC\,4191 & $9740^{+140} _{-150}$ & $12000^{+230} _{-130}$ & $12120^{+370} _{-410}$ & \cite{sharpee07}\\
Hen\,2-96 & $8990^{+170} _{-230}$ & - & $11540^{+800} _{-830}$ & \cite{garciarojas18}\\
M\,3-15 & $8360^{+250} _{-220}$ & - & $11210^{+660} _{-730}$ & \cite{garciarojas12}\\
Hen\,2-86 & $8320 \pm 210$ & $9450 \pm 370$ & $10260^{+1120} _{-780}$ & \cite{garciarojas12}\\
\botrule
\end{tabular*}
\end{minipage}
\end{center}
\end{table}

\setcounter{table}{5}
\begin{table}[h]
\begin{center}
\begin{minipage}{\textwidth}
\caption{Continued.}
\begin{tabular*}{\textwidth}{@{\extracolsep{\fill}}lcccccc@{\extracolsep{\fill}}}
\toprule%

Region & $T_{\rm e}$([O\,III])& $T_{\rm e}$([Ar\,III])& $T_{\rm e}$([N\,II]) &Ref.\\
 & K & K & K &    \\
\midrule

M\,2-31 & $9900^{+270} _{-190}$ & - & $11090^{+460} _{-430}$ & \cite{garciarojas18}\\
NGC\,5189 & $11450^{+300} _{-280}$ & $10310^{+460} _{-430}$ & $9480^{+450} _{-320}$ & \cite{garciarojas12}\\
M\,1-60 & $8800^{+210} _{-200}$ & - & $9810^{+540} _{-560}$ & \cite{garciarojas18}\\
Hen\,2-73 & $11770^{+380} _{-220}$ & $9650^{+1060} _{-1040}$ & $11480^{+460} _{-500}$ & \cite{garciarojas18}\\
IC\,4776 & $9960^{+150} _{-210}$ & $9230 \pm 190$ & $10110^{+860} _{-920}$ & \cite{sowicka17}\\
M\,1-25 & $7800 \pm 170$ & - & $8290^{+370} _{-280}$ & \cite{garciarojas12}\\
M\,1-33 & $8930^{+170} _{-200}$ & - & $9530^{+260} _{-270}$ & \cite{garciarojas18}\\
M\,2-36 & $8390^{+120} _{-100}$ & $7020^{+270} _{-590}$ & $14350^{+270} _{-480}$ & \cite{espiritu21}\\
Abell\,46 & $12930^{+260} _{-280}$ & - & $29300^{+18610} _{-9210}$ & \cite{corradi15}\\
\botrule
\end{tabular*}
\end{minipage}
\end{center}
\end{table}

\begin{table}[h]
\begin{center}
\begin{minipage}{\textwidth}
\caption{O\,II V1 fluxes and their associated physical parameters of the H\,II regions and other nebulae.}\label{tab:thingsofOII}
\begin{tabular*}{\textwidth}{@{\extracolsep{\fill}}lccccccccccccc@{\extracolsep{\fill}}}
\toprule%
Region & $T_{\rm e}$(O\,II/[O\,III])& $T_0$(O$^{2+}$) &$t^2$(O$^{2+}$) & O\,II V1/H$\beta$ & No. & Ref.\\
 & K & K & &  H$\beta$=100 & lines &   \\

\midrule
\multicolumn{7}{c}{{\bf Extragalactic H\,II regions}}\\
\\
NGC\,5461 & $7640 \pm 880$ & $7480 \pm 1080$ & $0.027 \pm 0.030$ & $0.387 \pm 0.123$ & 2 & \cite{esteban09}\\
NGC\,588 & $9320 \pm 600$ & $9330 \pm 740$ & $0.044 \pm 0.025$ & $0.281 \pm 0.057$ & 5 & \cite{Toribio16}\\
NGC\,595 & $6380 \pm 270$ & $6240 \pm 430$ & $0.031 \pm 0.015$ & $0.290 \pm 0.057$ & 2 & \cite{esteban09}\\
K\,932 & $7460 \pm 310$ & $7350 \pm 370$ & $0.026 \pm 0.011$ & $0.415 \pm 0.070$ & 5 & \cite{esteban09}\\
IC\,132 & $9340 \pm 730$ & $9390 \pm 830$ & $0.026 \pm 0.029$ & $0.323 \pm 0.078$ & 3 & \cite{Toribio16}\\
N\,11B & $8140 \pm 150$ & $8060 \pm 170$ & $0.031 \pm 0.006$ & $0.307 \pm 0.023$ & 5 & \cite{dominguezguzman22}\\
NGC\,5471 & $12700 \pm 2100$ & $12890 \pm 2210$ & $0.032 \pm 0.078$ & $0.146 \pm 0.050$ & 2 & \cite{esteban20}\\
N\,66A & $11230 \pm 440$ & $11330 \pm 450$ & $0.041 \pm 0.015$ & $0.128 \pm 0.015$ & 1 & \cite{dominguezguzman22}\\
NGC\,604 & $7270 \pm 250$ & $7170 \pm 320$ & $0.025 \pm 0.010$ & $0.354 \pm 0.055$ & 5 & \cite{esteban09}\\
NGC\,6822 & $9740 \pm 510$ & $9720 \pm 590$ & $0.055 \pm 0.018$ & $0.277 \pm 0.046$ & 6 & \cite{esteban14}\\
UV-1 & $8420 \pm 1080$ & $8390 \pm 1410$ & $0.071 \pm 0.037$ & $0.398 \pm 0.126$ & 2 & \cite{lopezsanchez07}\\
NGC\,5408 & $10610 \pm 900$ & $10890 \pm 1170$ & $0.162 \pm 0.032$ & $0.207 \pm 0.044$ & 6 & \cite{esteban14}\\
NGC\,1714 & $8470 \pm 410$ & $8410 \pm 490$ & $0.033 \pm 0.015$ & $0.368 \pm 0.064$ & 6 & \cite{dominguezguzman22}\\
N\,81 & $10100 \pm 180$ & $10140 \pm 190$ & $0.084 \pm 0.009$ & $0.216 \pm 0.013$ & 5 & \cite{dominguezguzman22}\\
IC\,2111 & $8300 \pm 530$ & $8230 \pm 620$ & $0.025 \pm 0.018$ & $0.285 \pm 0.060$ & 7 & \cite{dominguezguzman22}\\
NGC\,2363 & $12160 \pm 1330$ & $12730 \pm 1550$ & $0.129 \pm 0.048$ & $0.142 \pm 0.035$ & 5 & \cite{esteban09}\\
30Dor & $8970 \pm 270$ & $8940 \pm 300$ & $0.027 \pm 0.009$ & $0.339 \pm 0.037$ & 8 & \cite{peimbert03}\\
N\,44C & $9010 \pm 190$ & $8930 \pm 230$ & $0.070 \pm 0.007$ & $0.483 \pm 0.039$ & 6 & \cite{dominguezguzman22}\\
HII-2 & $8470 \pm 1050$ & $8130 \pm 1450$ & $0.108 \pm 0.036$ & $0.477 \pm 0.162$ & 1 & \cite{lopezsanchez07}\\
NGC\,5455 & $9000 \pm 1360$ & $9000 \pm 1520$ & $0.018 \pm 0.045$ & $0.251 \pm 0.083$ & 3 & \cite{esteban20}\\
VS44 & $7220 \pm 940$ & $7050 \pm 1200$ & $0.031 \pm 0.033$ & $0.332 \pm 0.120$ & 5 & \cite{esteban09}\\
N\,88A & $11860 \pm 180$ & $12060 \pm 190$ & $0.060 \pm 0.010$ & $0.183 \pm 0.008$ & 5 & \cite{dominguezguzman22}\\
HII-1 & $8540 \pm 1020$ & $8360 \pm 1460$ & $0.105 \pm 0.037$ & $0.477 \pm 0.161$ & 1 & \cite{lopezsanchez07}\\
\\
\multicolumn{7}{c}{{\bf Galactic ring nebulae}}\\
\\
NGC\,7635A2 & $6140 \pm 650$ & $5810 \pm 970$ & $0.060 \pm 0.028$ & $0.888 \pm 0.262$ & 3 & \cite{esteban16}\\
NGC\,7635A3 & $6280 \pm 430$ & $6040 \pm 670$ & $0.041 \pm 0.023$ & $0.695 \pm 0.189$ & 3 & \cite{esteban16}\\
NGC\,7635A4 & $7170 \pm 570$ & $7020 \pm 780$ & $0.045 \pm 0.022$ & $0.104 \pm 0.027$ & 2 & \cite{esteban16}\\
NGC\,6888A2 & $5060 \pm 170$ & $2380 \pm 650$ & $0.244 \pm 0.014$ & $2.224 \pm 0.434$ & 1 & \cite{esteban16}\\
NGC\,6888A3 & $5470 \pm 260$ & $4280 \pm 680$ & $0.145 \pm 0.022$ & $0.983 \pm 0.229$ & 3 & \cite{esteban16}\\
NGC\,6888A4 & $6770 \pm 770$ & $6180 \pm 1330$ & $0.104 \pm 0.039$ & $0.577 \pm 0.169$ & 4 & \cite{esteban16}\\
NGC\,6888A5 & $6470 \pm 410$ & $5840 \pm 730$ & $0.106 \pm 0.021$ & $0.721 \pm 0.195$ & 5 & \cite{esteban16}\\
NGC\,6888A6 & $6940 \pm 320$ & $6520 \pm 580$ & $0.084 \pm 0.024$ & $0.488 \pm 0.097$ & 3 & \cite{esteban16}\\
\botrule
\end{tabular*}
\end{minipage}
\end{center}
\end{table}

\setcounter{table}{6}
\begin{table}[h]
\begin{center}
\begin{minipage}{\textwidth}
\caption{Continued.}
\begin{tabular*}{\textwidth}{@{\extracolsep{\fill}}lccccccccccccc@{\extracolsep{\fill}}}
\toprule%
Region & $T_{\rm e}$(O\,II/[O\,III])& $T_0$(O$^{2+}$) &$t^2$(O$^{2+}$) & O\,II V1/H$\beta$ & No. & Ref.\\
 & K & K & &  H$\beta$=100 & lines &   \\

\midrule
\multicolumn{7}{c}{{\bf Galactic H\,II regions}}\\
\\
NGC\,3603 & $7550 \pm 560$ & $7410 \pm 720$ & $0.046 \pm 0.019$ & $0.644 \pm 0.167$ & 4 & \cite{garciarojas06}\\
M\,42-P1 & $7580 \pm 230$ & $7510 \pm 280$ & $0.022 \pm 0.008$ & $0.417 \pm 0.052$ & 6 & \cite{delgadoinglada16}\\
M\,42-2 & $7390 \pm 60$ & $7250 \pm 80$ & $0.035 \pm 0.003$ & $0.472 \pm 0.015$ & 6 & \cite{mendezdelgado22b}\\
Sh\,2-100 & $7380 \pm 320$ & $7280 \pm 400$ & $0.028 \pm 0.011$ & $0.501 \pm 0.091$ & 6 & \cite{esteban17}\\
M\,42-3 & $7660 \pm 120$ & $7560 \pm 140$ & $0.027 \pm 0.004$ & $0.403 \pm 0.025$ & 5 & \cite{mendezdelgado21a}\\
M\,42-1 & $7730 \pm 130$ & $7660 \pm 160$ & $0.022 \pm 0.005$ & $0.355 \pm 0.025$ & 5 & \cite{mendezdelgado21a}\\
Sh\,2-311 & $7770 \pm 900$ & $7750 \pm 1120$ & $0.035 \pm 0.031$ & $0.152 \pm 0.048$ & 5 & \cite{garciarojas05}\\
M\,16 & $6240 \pm 320$ & $6010 \pm 420$ & $0.042 \pm 0.011$ & $0.252 \pm 0.058$ & 5 & \cite{garciarojas06}\\
M\,42-bar & $7620 \pm 260$ & $7550 \pm 290$ & $0.026 \pm 0.008$ & $0.198 \pm 0.028$ & 6 & \cite{delgadoinglada16}\\
M\,42-1 & $7170 \pm 110$ & $7070 \pm 140$ & $0.026 \pm 0.004$ & $0.181 \pm 0.012$ & 5 & \cite{mendezdelgado21b}\\
NGC\,3576 & $7140 \pm 130$ & $6960 \pm 160$ & $0.041 \pm 0.004$ & $0.559 \pm 0.043$ & 6 & \cite{garciarojas04}\\
M\,42 & $7370 \pm 130$ & $7250 \pm 160$ & $0.031 \pm 0.005$ & $0.499 \pm 0.038$ & 6 & \cite{esteban04}\\
M\,17 & $6980 \pm 280$ & $6870 \pm 340$ & $0.029 \pm 0.009$ & $0.638 \pm 0.108$ & 6 & \cite{garciarojas07}\\
M\,8 & $6670 \pm 210$ & $6470 \pm 280$ & $0.043 \pm 0.008$ & $0.210 \pm 0.030$ & 6 & \cite{garciarojas07}\\
M\,42-4 & $7550 \pm 100$ & $7470 \pm 120$ & $0.024 \pm 0.004$ & $0.403 \pm 0.021$ & 5 & \cite{mendezdelgado21a}\\
M\,42-3 & $7320 \pm 80$ & $7200 \pm 100$ & $0.033 \pm 0.003$ & $0.498 \pm 0.022$ & 6 & \cite{mendezdelgado22b}\\
Sh\,2-288 & $7440 \pm 1080$ & $7420 \pm 1520$ & $0.053 \pm 0.042$ & $0.170 \pm 0.062$ & 2 & \cite{esteban17}\\
M\,42 & $7460 \pm 290$ & $7380 \pm 360$ & $0.022 \pm 0.011$ & $0.394 \pm 0.061$ & 6 & \cite{mesadelgado09}\\
M\,42-2 & $6710 \pm 150$ & $6480 \pm 180$ & $0.044 \pm 0.005$ & $0.305 \pm 0.030$ & 4 & \cite{mendezdelgado21b}\\
M\,42-2 & $7670 \pm 70$ & $7600 \pm 90$ & $0.022 \pm 0.003$ & $0.382 \pm 0.012$ & 6 & \cite{mendezdelgado21a}\\
\\
\multicolumn{7}{c}{{\bf Extragalactic H\,II regions from the literature}}\\
\\
NGC\,456a-1 & $9360 \pm 780$ & $9270 \pm 970$ & $0.089 \pm 0.026$ & $0.212 \pm 0.050$ & 2 & \cite{guseva11}\\
NGC\,456a-2 & $8920 \pm 1020$ & $8830 \pm 1330$ & $0.092 \pm 0.035$ & $0.294 \pm 0.095$ & 2 & \cite{guseva11}\\
NGC\,6822V & $10030 \pm 1090$ & $10040 \pm 1240$ & $0.051 \pm 0.037$ & $0.249 \pm 0.071$ & 4 & \cite{peimbert05}\\
NGC\,5253C1 & $11820 \pm 720$ & $11910 \pm 690$ & $0.009 \pm 0.023$ & $0.166 \pm 0.027$ & 2 & \cite{guseva11}\\
Mrk1259 & $5400 \pm 400$ & $4220 \pm 760$ & $0.143 \pm 0.018$ & $1.160 \pm 0.333$ & 2 & \cite{guseva11}\\
NGC\,5253C2 & $9290 \pm 590$ & $9270 \pm 650$ & $0.027 \pm 0.019$ & $0.270 \pm 0.056$ & 2 & \cite{guseva11}\\
NGC\,5253P2 & $10030 \pm 910$ & $10000 \pm 1070$ & $0.071 \pm 0.030$ & $0.281 \pm 0.065$ & 2 & \cite{guseva11}\\
NGC\,346 & $9840 \pm 530$ & $9830 \pm 630$ & $0.092 \pm 0.018$ & $0.261 \pm 0.045$ & 4 & \cite{valerdi19}\\
NGC\,456-2 & $9940 \pm 1230$ & $9970 \pm 1510$ & $0.065 \pm 0.043$ & $0.189 \pm 0.054$ & 4 & \cite{penaguerrero12}\\
\botrule
\end{tabular*}
\end{minipage}
\end{center}
\end{table}

\setcounter{table}{6}
\begin{table}[h]
\begin{center}
\begin{minipage}{\textwidth}
\caption{Continued.}
\begin{tabular*}{\textwidth}{@{\extracolsep{\fill}}lccccccccccccc@{\extracolsep{\fill}}}
\toprule%
Region & $T_{\rm e}$(O\,II/[O\,III])& $T_0$(O$^{2+}$) &$t^2$(O$^{2+}$) & O\,II V1/H$\beta$ & No. & Ref.\\
 & K & K & &  H$\beta$=100 & lines &   \\

\midrule
\multicolumn{7}{c}{{\bf Galactic planetary nebulae}}\\
\\
NGC\,2440 & $11470 \pm 330$ & $11750 \pm 370$ & $0.104 \pm 0.014$ & $0.394 \pm 0.031$ & 4 & \cite{sharpee07}\\
Hb\,4 & $7030 \pm 270$ & $6600 \pm 400$ & $0.093 \pm 0.012$ & $1.808 \pm 0.282$ & 4 & \cite{garciarojas12}\\
NGC\,5315 & $7530 \pm 230$ & $7390 \pm 300$ & $0.038 \pm 0.011$ & $1.011 \pm 0.106$ & 6 & \cite{madonna17}\\
M\,1-31 & $6970 \pm 550$ & $6820 \pm 730$ & $0.041 \pm 0.020$ & $1.300 \pm 0.309$ & 3 & \cite{garciarojas18}\\
Cn\,1-5 & $7370 \pm 450$ & $7180 \pm 580$ & $0.043 \pm 0.016$ & $1.138 \pm 0.251$ & 3 & \cite{garciarojas12}\\
H\,1-40 & $7600 \pm 1040$ & $7360 \pm 1440$ & $0.077 \pm 0.038$ & $1.084 \pm 0.421$ & 1 & \cite{garciarojas18}\\
H\,1-50 & $8290 \pm 390$ & $8070 \pm 500$ & $0.086 \pm 0.015$ & $1.259 \pm 0.204$ & 3 & \cite{garciarojas18}\\
Hen\,2158 & $8190 \pm 1010$ & $8200 \pm 1220$ & $0.032 \pm 0.034$ & $0.441 \pm 0.159$ & 2 & \cite{garciarojas18}\\
IC\,418 & $7910 \pm 230$ & $7830 \pm 260$ & $0.027 \pm 0.008$ & $0.207 \pm 0.023$ & 8 & \cite{sharpee03}\\
PC\,14 & $7630 \pm 400$ & $7460 \pm 510$ & $0.050 \pm 0.014$ & $1.362 \pm 0.264$ & 3 & \cite{garciarojas12}\\
M\,1-32 & $7050 \pm 260$ & $6710 \pm 390$ & $0.072 \pm 0.011$ & $0.691 \pm 0.105$ & 2 & \cite{garciarojas12}\\
M\,1-61 & $7760 \pm 350$ & $7620 \pm 440$ & $0.045 \pm 0.013$ & $0.946 \pm 0.162$ & 5 & \cite{garciarojas12}\\
Ou5 & $4380 \pm 130$ & $1060 \pm 630$ & $0.258 \pm 0.014$ & $14.436 \pm 2.766$ & 4 & \cite{corradi15}\\
NGC\,6369 & $9220 \pm 630$ & $9180 \pm 690$ & $0.044 \pm 0.022$ & $0.624 \pm 0.133$ & 3 & \cite{garciarojas12}\\
Pe\,1-1 & $8270 \pm 550$ & $8130 \pm 700$ & $0.057 \pm 0.020$ & $0.815 \pm 0.187$ & 4 & \cite{garciarojas12}\\
IC\,2501 & $9270 \pm 180$ & $9260 \pm 190$ & $0.005 \pm 0.006$ & $0.463 \pm 0.030$ & 8 & \cite{sharpee07}\\
M\,1-30 & $5440 \pm 120$ & $5180 \pm 190$ & $0.038 \pm 0.006$ & $1.032 \pm 0.116$ & 5 & \cite{garciarojas12}\\
IC\,4191 & $8040 \pm 100$ & $7900 \pm 130$ & $0.052 \pm 0.005$ & $1.460 \pm 0.062$ & 8 & \cite{sharpee07}\\
Hen\,2-96 & $8000 \pm 590$ & $7890 \pm 710$ & $0.032 \pm 0.020$ & $0.780 \pm 0.184$ & 5 & \cite{garciarojas18}\\
M\,3-15 & $6720 \pm 660$ & $6530 \pm 910$ & $0.049 \pm 0.024$ & $1.792 \pm 0.504$ & 2 & \cite{garciarojas12}\\
Hen\,2-86 & $7060 \pm 160$ & $6890 \pm 220$ & $0.039 \pm 0.008$ & $1.321 \pm 0.116$ & 7 & \cite{garciarojas12}\\
M\,2-31 & $7900 \pm 530$ & $7680 \pm 660$ & $0.063 \pm 0.018$ & $1.091 \pm 0.260$ & 4 & \cite{garciarojas18}\\
NGC\,5189 & $9370 \pm 780$ & $9300 \pm 960$ & $0.065 \pm 0.028$ & $0.632 \pm 0.152$ & 5 & \cite{garciarojas12}\\
M\,1-60 & $7010 \pm 250$ & $6770 \pm 360$ & $0.056 \pm 0.010$ & $1.704 \pm 0.247$ & 4 & \cite{garciarojas18}\\
Hen\,2-73 & $8770 \pm 610$ & $8570 \pm 800$ & $0.094 \pm 0.022$ & $0.998 \pm 0.228$ & 4 & \cite{garciarojas18}\\
IC\,4776 & $8390 \pm 120$ & $8280 \pm 150$ & $0.048 \pm 0.007$ & $0.733 \pm 0.034$ & 7 & \cite{sowicka17}\\
M\,1-25 & $6640 \pm 340$ & $6450 \pm 460$ & $0.037 \pm 0.013$ & $0.994 \pm 0.213$ & 4 & \cite{garciarojas12}\\
M\,1-33 & $6780 \pm 210$ & $6460 \pm 300$ & $0.067 \pm 0.009$ & $1.931 \pm 0.249$ & 4 & \cite{garciarojas18}\\
M\,2-36 & $5450 \pm 70$ & $4770 \pm 140$ & $0.092 \pm 0.004$ & $4.826 \pm 0.374$ & 8 & \cite{espiritu21}\\
Abell\,46 & $4240 \pm 60$ & $320 \pm 370$ & $0.279 \pm 0.008$ & $11.149 \pm 1.056$ & 6 & \cite{corradi15}\\
\botrule
\end{tabular*}
\end{minipage}
\end{center}
\end{table}

\begin{table}[h]
\begin{center}
\begin{minipage}{\textwidth}
\caption{Ionic abundances of the H\,II regions and other nebulae.}\label{tab:ionicabundances}
\begin{tabular*}{\textwidth}{@{\extracolsep{\fill}}lcccccc@{\extracolsep{\fill}}}
\toprule%

Region & 12+log(O$^{+}$/H$^{+}$)&12+log(O$^{2+}$/H$^{+}$)&12+log(O$^{2+}$/H$^{+}$)& Ref.\\
 & CELs & CELs $t^2$(O$^{2+}$)=0 & RLs &  &   \\

\midrule
\multicolumn{5}{c}{{\bf Extragalactic H\,II regions}}\\
\\
NGC\,5461 & $7.78^{+0.08} _{-0.06}$ & $8.29 \pm 0.05$ & $8.46^{+0.14} _{-0.13}$ & \cite{esteban09}\\
NGC\,588 & $7.40^{+0.27} _{-0.11}$ & $8.11^{+0.09} _{-0.06}$ & $8.33 \pm 0.08$ & \cite{Toribio16}\\
NGC\,595 & $8.29^{+0.05} _{-0.04}$ & $8.02^{+0.14} _{-0.05}$ & $8.34^{+0.08} _{-0.09}$ & \cite{esteban09}\\
K\,932 & $7.87^{+0.05} _{-0.03}$ & $8.30 \pm 0.03$ & $8.50 \pm 0.07$ & \cite{esteban09}\\
IC\,132 & $7.17^{+0.21} _{-0.14}$ & $8.24^{+0.09} _{-0.06}$ & $8.38 \pm 0.10$ & \cite{Toribio16}\\
N\,11B & $7.99^{+0.03} _{-0.04}$ & $8.17^{+0.03} _{-0.01}$ & $8.37 \pm 0.03$ & \cite{dominguezguzman22}\\
NGC\,5471 & $7.17^{+0.07} _{-0.06}$ & $7.93^{+0.02} _{-0.01}$ & $8.05 \pm 0.15$ & \cite{esteban20}\\
N\,66A & $7.49^{+0.06} _{-0.04}$ & $7.84 \pm 0.02$ & $7.99 \pm 0.05$ & \cite{dominguezguzman22}\\
NGC\,604 & $7.92 \pm 0.06$ & $8.23 \pm 0.04$ & $8.43 \pm 0.07$ & \cite{esteban09}\\
NGC\,6822 & $7.37^{+0.36} _{-0.16}$ & $8.08^{+0.03} _{-0.02}$ & $8.31^{+0.08} _{-0.07}$ & \cite{esteban14}\\
UV-1 & $7.80^{+0.14} _{-0.09}$ & $8.09 \pm 0.02$ & $8.50^{+0.13} _{-0.14}$ & \cite{lopezsanchez07}\\
NGC\,5408 & $7.29^{+0.14} _{-0.12}$ & $7.70 \pm 0.03$ & $8.18 \pm 0.10$ & \cite{esteban14}\\
NGC\,1714 & $7.65^{+0.13} _{-0.08}$ & $8.27^{+0.03} _{-0.05}$ & $8.45 \pm 0.07$ & \cite{dominguezguzman22}\\
N\,81 & $7.30^{+0.04} _{-0.03}$ & $7.90^{+0.03} _{-0.02}$ & $8.22 \pm 0.03$ & \cite{dominguezguzman22}\\
IC\,2111 & $8.05^{+0.08} _{-0.06}$ & $8.18 \pm 0.04$ & $8.34^{+0.09} _{-0.10}$ & \cite{dominguezguzman22}\\
NGC\,2363 & $6.57^{+0.12} _{-0.13}$ & $7.69^{+0.02} _{-0.03}$ & $8.02^{+0.11} _{-0.10}$ & \cite{esteban09}\\
30Dor & $7.62 \pm 0.05$ & $8.27 \pm 0.01$ & $8.42 \pm 0.05$ & \cite{peimbert03}\\
N\,44C & $7.30 \pm 0.05$ & $8.23 \pm 0.02$ & $8.56^{+0.04} _{-0.03}$ & \cite{dominguezguzman22}\\
HII-2 & $7.66^{+0.11} _{-0.10}$ & $8.06 \pm 0.02$ & $8.57^{+0.15} _{-0.14}$ & \cite{lopezsanchez07}\\
NGC\,5455 & $7.74 \pm 0.03$ & $8.18 \pm 0.03$ & $8.28^{+0.15} _{-0.14}$ & \cite{esteban20}\\
VS44 & $7.91^{+0.08} _{-0.06}$ & $8.15^{+0.04} _{-0.06}$ & $8.41 \pm 0.15$ & \cite{esteban09}\\
N\,88A & $6.88 \pm 0.07$ & $8.00^{+0.02} _{-0.01}$ & $8.18 \pm 0.02$ & \cite{dominguezguzman22}\\
HII-1 & $7.59^{+0.09} _{-0.06}$ & $8.07 \pm 0.02$ & $8.57 \pm 0.16$ & \cite{lopezsanchez07}\\
\\
\multicolumn{5}{c}{{\bf Galactic ring nebulae}}\\
\\
NGC\,7635A2 & $7.93^{+0.31} _{-0.11}$ & $8.24^{+0.11} _{-0.08}$ & $8.82 \pm 0.13$ & \cite{esteban16}\\
NGC\,7635A3 & $8.06^{+0.16} _{-0.07}$ & $8.31^{+0.18} _{-0.10}$ & $8.73 \pm 0.11$ & \cite{esteban16}\\
NGC\,7635A4 & $8.16^{+0.03} _{-0.05}$ & $7.60^{+0.10} _{-0.08}$ & $7.95 \pm 0.11$ & \cite{esteban16}\\
NGC\,6888A2 & $8.33 \pm 0.04$ & $7.52 \pm 0.03$ & $9.23 \pm 0.09$ & \cite{esteban16}\\
NGC\,6888A3 & $8.13 \pm 0.06$ & $7.69^{+0.15} _{-0.09}$ & $8.86 \pm 0.10$ & \cite{esteban16}\\
NGC\,6888A4 & $7.47^{+0.06} _{-0.05}$ & $7.95 \pm 0.13$ & $8.64^{+0.12} _{-0.13}$ & \cite{esteban16}\\
NGC\,6888A5 & $7.49 \pm 0.06$ & $7.97^{+0.09} _{-0.07}$ & $8.74^{+0.12} _{-0.11}$ & \cite{esteban16}\\
NGC\,6888A6 & $7.57^{+0.05} _{-0.06}$ & $7.98^{+0.17} _{-0.11}$ & $8.56 \pm 0.09$ & \cite{esteban16}\\
\botrule
\end{tabular*}
\end{minipage}
\end{center}
\end{table}

\setcounter{table}{7}
\begin{table}[h]
\begin{center}
\begin{minipage}{\textwidth}
\caption{Continued.}
\begin{tabular*}{\textwidth}{@{\extracolsep{\fill}}lcccccc@{\extracolsep{\fill}}}
\toprule%

Region & 12+log(O$^{+}$/H$^{+}$)&12+log(O$^{2+}$/H$^{+}$)&12+log(O$^{2+}$/H$^{+}$)& Ref.\\
 & CELs & CELs $t^2$(O$^{2+}$)=0 & RLs &  &   \\
\midrule
\multicolumn{5}{c}{{\bf Galactic H\,II regions}}\\
\\
NGC\,3603 & $7.36^{+0.13} _{-0.09}$ & $8.43^{+0.02} _{-0.03}$ & $8.74 \pm 0.11$ & \cite{garciarojas06}\\
M\,42-P1 & $7.81^{+0.10} _{-0.07}$ & $8.39 \pm 0.02$ & $8.55 \pm 0.05$ & \cite{delgadoinglada16}\\
M\,42-2 & $7.74^{+0.07} _{-0.05}$ & $8.36 \pm 0.02$ & $8.61^{+0.02} _{-0.01}$ & \cite{mendezdelgado22b}\\
Sh\,2-100 & $7.76^{+0.11} _{-0.06}$ & $8.38 \pm 0.02$ & $8.59 \pm 0.08$ & \cite{esteban17}\\
M\,42-3 & $7.75^{+0.07} _{-0.06}$ & $8.36^{+0.01} _{-0.02}$ & $8.55 \pm 0.03$ & \cite{mendezdelgado21a}\\
M\,42-1 & $7.87 \pm 0.06$ & $8.33^{+0.02} _{-0.01}$ & $8.49 \pm 0.03$ & \cite{mendezdelgado21a}\\
Sh\,2-311 & $8.30 \pm 0.04$ & $7.82 \pm 0.02$ & $8.05^{+0.14} _{-0.15}$ & \cite{garciarojas05}\\
M\,16 & $8.47^{+0.06} _{-0.05}$ & $7.92^{+0.04} _{-0.03}$ & $8.31^{+0.10} _{-0.11}$ & \cite{garciarojas06}\\
M\,42-bar & $8.39 \pm 0.04$ & $8.04 \pm 0.02$ & $8.23 \pm 0.06$ & \cite{delgadoinglada16}\\
M\,42-1 & $8.14^{+0.03} _{-0.04}$ & $7.96 \pm 0.01$ & $8.17 \pm 0.03$ & \cite{mendezdelgado21b}\\
NGC\,3576 & $8.06^{+0.08} _{-0.05}$ & $8.36 \pm 0.01$ & $8.66 \pm 0.03$ & \cite{garciarojas04}\\
M\,42 & $7.81^{+0.07} _{-0.06}$ & $8.40 \pm 0.01$ & $8.64 \pm 0.03$ & \cite{esteban04}\\
M\,17 & $7.82^{+0.06} _{-0.04}$ & $8.45 \pm 0.02$ & $8.69^{+0.08} _{-0.07}$ & \cite{garciarojas07}\\
M\,8 & $8.34^{+0.04} _{-0.03}$ & $7.88^{+0.02} _{-0.03}$ & $8.24 \pm 0.06$ & \cite{garciarojas07}\\
M\,42-4 & $7.81^{+0.06} _{-0.05}$ & $8.36 \pm 0.02$ & $8.54 \pm 0.02$ & \cite{mendezdelgado21a}\\
M\,42-3 & $7.68 \pm 0.05$ & $8.38^{+0.01} _{-0.02}$ & $8.63 \pm 0.02$ & \cite{mendezdelgado22b}\\
Sh\,2-288 & $8.22^{+0.08} _{-0.07}$ & $7.72^{+0.11} _{-0.08}$ & $8.11^{+0.17} _{-0.15}$ & \cite{esteban17}\\
M\,42 & $7.94 \pm 0.07$ & $8.35^{+0.05} _{-0.04}$ & $8.51 \pm 0.07$ & \cite{mesadelgado09}\\
M\,42-2 & $8.17^{+0.03} _{-0.04}$ & $8.03^{+0.02} _{-0.01}$ & $8.40 \pm 0.04$ & \cite{mendezdelgado21b}\\
M\,42-2 & $7.82 \pm 0.05$ & $8.36 \pm 0.02$ & $8.52 \pm 0.01$ & \cite{mendezdelgado21a}\\
\\
\multicolumn{5}{c}{{\bf Extragalactic H\,II regions from the literature}}\\
\\
NGC\,456a-1 & $7.65 \pm 0.06$ & $7.84 \pm 0.01$ & $8.22 \pm 0.10$ & \cite{guseva11}\\
NGC\,456a-2 & $7.53^{+0.25} _{-0.10}$ & $7.93 \pm 0.01$ & $8.34^{+0.15} _{-0.14}$ & \cite{guseva11}\\
NGC\,6822V & $6.91^{+0.25} _{-0.14}$ & $8.06^{+0.03} _{-0.02}$ & $8.27^{+0.13} _{-0.14}$ & \cite{peimbert05}\\
NGC\,5253C1 & $7.34 \pm 0.02$ & $8.08 \pm 0.01$ & $8.11 \pm 0.07$ & \cite{guseva11}\\
Mrk1259 & $8.36^{+0.08} _{-0.07}$ & $7.75^{+0.06} _{-0.05}$ & $8.97 \pm 0.13$ & \cite{guseva11}\\
NGC\,5253C2 & $7.56^{+0.07} _{-0.05}$ & $8.17 \pm 0.01$ & $8.31 \pm 0.09$ & \cite{guseva11}\\
NGC\,5253P2 & $7.42^{+0.04} _{-0.03}$ & $8.07 \pm 0.01$ & $8.35 \pm 0.10$ & \cite{guseva11}\\
NGC\,346 & $7.32^{+0.14} _{-0.11}$ & $7.93 \pm 0.01$ & $8.28^{+0.08} _{-0.07}$ & \cite{valerdi19}\\
NGC\,456-2 & $7.54^{+0.15} _{-0.11}$ & $7.88^{+0.02} _{-0.01}$ & $8.16^{+0.12} _{-0.13}$ & \cite{penaguerrero12}\\
\botrule
\end{tabular*}
\end{minipage}
\end{center}
\end{table}

\setcounter{table}{7}
\begin{table}[h]
\begin{center}
\begin{minipage}{\textwidth}
\caption{Continued.}
\begin{tabular*}{\textwidth}{@{\extracolsep{\fill}}lcccccc@{\extracolsep{\fill}}}
\toprule%

Region & 12+log(O$^{+}$/H$^{+}$)&12+log(O$^{2+}$/H$^{+}$)&12+log(O$^{2+}$/H$^{+}$)& Ref.\\
 & CELs & CELs $t^2$(O$^{2+}$)=0 & RLs &  &   \\
\midrule
\multicolumn{5}{c}{{\bf Galactic planetary nebulae}}\\
\\
NGC\,2440 & $7.41^{+0.09} _{-0.06}$ & $8.22^{+0.02} _{-0.03}$ & $8.50^{+0.03} _{-0.04}$ & \cite{sharpee07}\\
Hb\,4 & $7.33^{+0.19} _{-0.10}$ & $8.59 \pm 0.05$ & $9.20 \pm 0.07$ & \cite{garciarojas12}\\
NGC\,5315 & $7.83^{+0.32} _{-0.23}$ & $8.69^{+0.07} _{-0.05}$ & $8.95 \pm 0.05$ & \cite{madonna17}\\
M\,1-31 & $7.68^{+0.13} _{-0.11}$ & $8.74^{+0.05} _{-0.04}$ & $9.06 \pm 0.10$ & \cite{garciarojas18}\\
Cn\,1-5 & $8.10^{+0.13} _{-0.07}$ & $8.69 \pm 0.04$ & $9.00^{+0.09} _{-0.10}$ & \cite{garciarojas12}\\
H\,1-40 & $7.02^{+0.13} _{-0.10}$ & $8.49^{+0.05} _{-0.04}$ & $8.96 \pm 0.17$ & \cite{garciarojas18}\\
H\,1-50 & $7.38^{+0.10} _{-0.07}$ & $8.60^{+0.03} _{-0.04}$ & $9.04 \pm 0.07$ & \cite{garciarojas18}\\
Hen\,2158 & $7.88 \pm 0.07$ & $8.35 \pm 0.03$ & $8.57^{+0.15} _{-0.16}$ & \cite{garciarojas18}\\
IC\,418 & $8.37^{+0.08} _{-0.06}$ & $8.08 \pm 0.02$ & $8.26^{+0.05} _{-0.04}$ & \cite{sharpee03}\\
PC\,14 & $7.35^{+0.11} _{-0.08}$ & $8.73^{+0.04} _{-0.03}$ & $9.07^{+0.08} _{-0.09}$ & \cite{garciarojas12}\\
M\,1-32 & $8.25^{+0.17} _{-0.11}$ & $8.28^{+0.05} _{-0.04}$ & $8.78^{+0.07} _{-0.06}$ & \cite{garciarojas12}\\
M\,1-61 & $7.15^{+0.24} _{-0.14}$ & $8.63^{+0.06} _{-0.05}$ & $8.93 \pm 0.07$ & \cite{garciarojas12}\\
Ou5 & $6.27^{+0.48} _{-0.17}$ & $7.98^{+0.03} _{-0.04}$ & $10.06^{+0.09} _{-0.08}$ & \cite{corradi15}\\
NGC\,6369 & $6.91^{+0.10} _{-0.07}$ & $8.50^{+0.05} _{-0.03}$ & $8.73 \pm 0.09$ & \cite{garciarojas12}\\
Pe\,1-1 & $7.48^{+0.25} _{-0.18}$ & $8.53^{+0.05} _{-0.04}$ & $8.85 \pm 0.10$ & \cite{garciarojas12}\\
IC\,2501 & $7.50^{+0.04} _{-0.05}$ & $8.58 \pm 0.02$ & $8.60 \pm 0.03$ & \cite{sharpee07}\\
M\,1-30 & $8.42^{+0.14} _{-0.08}$ & $8.50^{+0.06} _{-0.04}$ & $8.96^{+0.04} _{-0.05}$ & \cite{garciarojas12}\\
IC\,4191 & $7.30 \pm 0.08$ & $8.79^{+0.03} _{-0.02}$ & $9.10 \pm 0.02$ & \cite{sharpee07}\\
Hen\,2-96 & $7.43^{+0.16} _{-0.14}$ & $8.63^{+0.05} _{-0.04}$ & $8.83 \pm 0.10$ & \cite{garciarojas18}\\
M\,3-15 & $6.88^{+0.16} _{-0.08}$ & $8.78 \pm 0.05$ & $9.19^{+0.12} _{-0.11}$ & \cite{garciarojas12}\\
Hen\,2-86 & $7.42^{+0.30} _{-0.21}$ & $8.76^{+0.05} _{-0.04}$ & $9.07 \pm 0.04$ & \cite{garciarojas12}\\
M\,2-31 & $7.43 \pm 0.08$ & $8.60 \pm 0.05$ & $8.97 \pm 0.10$ & \cite{garciarojas18}\\
NGC\,5189 & $8.13^{+0.11} _{-0.07}$ & $8.43 \pm 0.04$ & $8.70^{+0.11} _{-0.10}$ & \cite{garciarojas12}\\
M\,1-60 & $7.74^{+0.21} _{-0.13}$ & $8.76^{+0.05} _{-0.04}$ & $9.17^{+0.06} _{-0.07}$ & \cite{garciarojas18}\\
Hen\,2-73 & $7.54^{+0.11} _{-0.10}$ & $8.51 \pm 0.05$ & $8.93 \pm 0.10$ & \cite{garciarojas18}\\
IC\,4776 & $7.76^{+0.45} _{-0.14}$ & $8.53 \pm 0.03$ & $8.81 \pm 0.02$ & \cite{sowicka17}\\
M\,1-25 & $8.29^{+0.10} _{-0.09}$ & $8.62 \pm 0.05$ & $8.94 \pm 0.09$ & \cite{garciarojas12}\\
M\,1-33 & $7.68^{+0.08} _{-0.05}$ & $8.72 \pm 0.04$ & $9.22 \pm 0.06$ & \cite{garciarojas18}\\
M\,2-36 & $6.95^{+0.05} _{-0.04}$ & $8.72^{+0.03} _{-0.02}$ & $9.61 \pm 0.03$ & \cite{espiritu21}\\
Abell\,46 & $6.25^{+1.05} _{-0.19}$ & $7.75^{+0.03} _{-0.02}$ & $9.95 \pm 0.04$ & \cite{corradi15}\\
\botrule
\end{tabular*}
\end{minipage}
\end{center}
\end{table}




\end{appendices}


\bibliography{sn-bibliography}


\end{document}